\documentclass[5p,times]{elsarticle}
\usepackage{hyperref}
\usepackage{graphicx}
\usepackage[section]{placeins}
\usepackage{amsmath,amssymb}
\biboptions{sort&compress}

\RequirePackage{amsthm,amsmath,amsfonts,amssymb}
\RequirePackage{multirow}
\RequirePackage{booktabs}
\RequirePackage{url}
\biboptions{authoryear,round}

\theoremstyle{plain}

\theoremstyle{remark}
\newtheorem{remark}{Remark}[section]

\begin{document}

\begin{frontmatter}

\title{Bayesian Inference Procedures for A/B Testing: An Overview}

\author[spotify]{M\aa rten Schultzberg}
\ead{mschultzberg@spotify.com}

\author[spotify]{Mattias Fr\aa nberg}

\affiliation[spotify]{organization={Experimentation Platform team, Confidence, Spotify},
  city={Stockholm},
  country={Sweden}}

\begin{abstract}
Bayesian inference for A/B testing is a family of prior and stopping-rule
configurations with fundamentally different statistical properties, but it is often discussed as a single method, and no systematic overview exists. This paper organizes common configurations
into a three-tier hierarchy: 1) posterior coherence with no error control,
2) false positive rates bounded under continuous monitoring via Bayes factor stopping, and 3) false discovery rate control and calibrated shrinkage via empirical Bayes. Many commercial platforms operate at the lowest tier by default. We show that Bayes factor stopping is near-optimal for a broad class of cost functions, including most proposed in the A/B testing literature; because the same rule also controls the false positive rate, the choice between a decision-theoretic and a frequentist formulation is largely one of parameterization. Furthermore, the empirical Bayes
prior is the only path to the third tier, but winner-selected corpora, pooled programs, and heterogeneous metrics can
each prevent calibration regardless of corpus size. Simulations against group-sequential and always-valid frequentist baselines show that flat-prior posterior stopping exactly reproduces naive peeking, that a well-calibrated empirical Bayes prior achieves the lowest estimation error, and that expected-loss stopping minimizes regret only when shipping a null-effect variant is nearly free. Error rates, estimation accuracy, and regret are all different risks, and the appropriate method follows from the risks an experimentation program needs to control, not the other way around.
\end{abstract}


\end{frontmatter}

\section{Introduction}
\label{sec:intro}

Every product decision made from an A/B test carries risk. Shipping a
harmful variant, missing a beneficial one, and accumulating incorrect
decisions across a program are all forms of risk that experimentation
is meant to manage. A Bayesian experimenter can manage these risks in
different ways, optimizing for expected loss per decision, for
estimation accuracy across a program, or for false positive and false
discovery rates. These are different operating characteristics, and each implies a
different optimal configuration of priors and stopping rules. Which
configuration is appropriate depends on which risks the program needs
to control.

Yet Bayesian inference for A/B testing is often discussed as if it
were a single method. It is a family of configurations with
fundamentally different statistical properties. Depending on the
choice of prior and stopping rule, a Bayesian A/B test can produce
results numerically identical to a frequentist z-test, maintain
unconditional Type~I error control under continuous monitoring
\citep{johari2022,ville1939}, minimize the expected risk of a launch
decision \citep{arrow1949,wald1948}, or calibrate false discovery
rates and shrink effect estimates toward realistic magnitudes
\citep{deng2015,deng2016}. Much of the public discourse around
Bayesian experimentation conflates these configurations, leading to
claims that are true for some configurations and false for others.
Despite growing adoption in experimentation platforms
\citep{wagenmakers2007,stucchio2015,deng2015,kamalbasha2021,lu2023},
no systematic overview of these configurations, their guarantees, and
their relationship to each other and to common frequentist methods is
available.

This paper aims to make it easier to compare the most common Bayesian
configurations, both within the Bayesian paradigm and against
frequentist alternatives. We organize the configurations into a
three-tier guarantee hierarchy ranging from posterior coherence (valid
belief updates with no error control) through bounded FPR via Bayes
factor stopping to per-metric FDR calibration and calibrated shrinkage
via empirical Bayes. Many existing Bayesian A/B testing systems by default
operate at the lowest tier
\citep[e.g.,][accessed 2025]{statsig2024,eppo2024}. For cost functions discussed in the A/B testing
literature whose terminal payoffs are affine in the posterior,
BF stopping is often near-optimal for the sequential launch
decision \citep{arrow1949,wald1948}, and the choice between a
decision-theoretic and a frequentist formulation of the stopping rule
is largely a choice of parameterization. Cost functions with nonlinear
posterior functionals, such as expected opportunity loss, fall outside
this class. The highest tier requires an
empirical Bayes prior estimated from a corpus of historical
experiments, and specific corpus failure modes can prevent calibration
regardless of how much data is collected.

\section{Setup}
\label{sec:model}

Consider an experiment with two groups, control $C$ and treatment $T$,
each receiving $n$ observations. The data accrue sequentially, and a
stopping rule decides at each observation whether to continue or stop.
The estimand is the average treatment effect
$\delta = \mu_T - \mu_C$, and the question is whether $\delta > 0$.
The platform-facing labels $A$ (control) and $B$ (treatment) are used
in decision criteria such as ``$P(B > A) > 0.95$.'' The maximum
likelihood estimator $\hat{\delta} = \bar{Y}_T - \bar{Y}_C$ satisfies
$\hat{\delta} \sim N(\delta,\, 2\sigma^2/n)$ by the central limit
theorem in large samples, with effective sample size $N_E = n/2$ and
standardized statistic $Z_n = \hat{\delta}\sqrt{N_E}/\sigma$.
Throughout the paper $\sigma^2$ is treated as known. In practice it is
estimated, and the key results (Ville's inequality, the BF closed form)
extend asymptotically to the estimated-variance case under standard
regularity conditions \citep{johari2022,vandervaart1998}.

Under a prior $\pi(\delta)$, the posterior is
$\pi(\delta \mid \text{data}) \propto L(\delta \mid
\text{data})\,\pi(\delta)$. Under a flat prior,
conjugate normal updating gives
\begin{equation}
  \delta \mid \text{data} \sim N\!\left(\hat{\delta},\; \sigma^2/N_E\right).
\end{equation}

The rest of this section describes three families of stopping rules
that use this model: posterior-based rules, Bayes factor rules, and
decision-theoretic rules.

\subsection{Posterior inference and stopping rules}
\label{sub:posterior_stopping}

The posterior can be summarized in many ways for inference. Three
summaries are common. The posterior mean
$E[\delta \mid \text{data}]$ serves as a point estimate (under a flat
prior it equals the MLE $\hat\delta$). Under an informative prior the
posterior mean is pulled toward the prior mean relative to the MLE.
This shrinkage is a general property of Bayesian updating under
proper priors and is present at every tier of the hierarchy. The
credible interval is constructed from the percentiles of the posterior
distribution. The posterior probability that the treatment is better
than control, $P(\delta > 0 \mid \text{data})$, is also written
$P(B > A)$.

Stopping rules based on these summaries stop when the posterior
probability exceeds a threshold, $P(B > A) > \gamma$ for some chosen
$\gamma$, or when the credible interval excludes zero.

\subsection{Bayes factor inference and stopping rules}
\label{sub:bf_stopping}

Stopping rules based on Bayes factors are common.
The Bayes factor $\mathrm{BF}_{10} = p(\text{data} \mid H_1) /
p(\text{data} \mid H_0)$ compares the marginal probability of the data
under two competing hypotheses \citep{kass1995}, where
$p(\text{data} \mid H_k) = \int p(\text{data} \mid \delta)\,
d\pi_k(\delta)$ requires a proper prior under $H_1$. A large
$\mathrm{BF}_{10}$ is evidence for $H_1$. The inverse
$\mathrm{BF}_{01} = 1/\mathrm{BF}_{10}$ quantifies evidence for
$H_0$. Unlike a frequentist non-significant result, which is merely inconclusive,
a $\mathrm{BF}_{01} > 3$ provides positive evidence that the treatment has
no effect \citep{kass1995,dienes2014}. The most common BF stopping
rule rejects $H_0$ when $\mathrm{BF}_{10}$ exceeds a threshold.

Under the normal-normal conjugate model with prior
$\delta \sim N(0, V^2)$, the BF has a closed form:
\begin{equation}
  \label{eq:bf_closed}
  \mathrm{BF}_{10} = \left(1 + \frac{N_E V^2}{\sigma^2}\right)^{-1/2}
  \cdot \exp\!\left(\frac{Z_n^2\,N_E V^2}
  {2\,\sigma^2\!\left(1 + N_E V^2/\sigma^2\right)}\right).
\end{equation}
We write $V$ for the prior standard deviation throughout.

\subsection{Decision-theoretic stopping rules}
\label{sub:decision_theoretic}

A decision-theoretic stopping rule explicitly models the costs and
benefits of making correct and incorrect decisions, and finds the policy
that minimizes the expected loss \citep{berger1985}. Different loss
functions lead to different stopping rules.

\textbf{Expected opportunity loss.} The rule of \citet{stucchio2015}
defines loss as the expected cost of choosing the suboptimal treatment,
measured in the metric's units. The rule stops when the expected regret
of the better of the two possible decisions (the minimum) falls below a tolerance $\varepsilon$:
\begin{align}
\min\bigl(E[\max(\delta, 0)],\, E[\max(-\delta, 0)]\bigr) < \varepsilon.
\end{align}
This formulation of expected regret as an A/B test stopping rule was
introduced in a practitioner whitepaper for the VWO platform
($\varepsilon = 0.02$ by default) and adopted by several industry
teams \citep{stucchio2015,frasco2018}. The stopping criterion depends
on the posterior, so this is a posterior-based stopping rule with a
decision-theoretic motivation. When development costs are already sunk and the
only post-launch risk is deploying a harmful variant, minimizing
expected loss in the metric's units is a natural objective.

\textbf{Misclassification costs.} An alternative loss assigns a
per-observation sampling cost $c > 0$, a false-positive cost
$K_I > 0$, and a false-negative cost $K_{II} > 0$:
\begin{equation}
  \label{eq:loss_main}
  \ell(\tau, D_\tau, M) = c\,\tau
    + K_I\,\mathbf{1}\{D_\tau{=}1,\,M{=}H_0\}
    + K_{II}\,\mathbf{1}\{D_\tau{=}0,\,M{=}H_1\}.
\end{equation}
Notation is defined in \ref{app:bellman}.

Minimizing this loss over all sequential policies (\ref{app:bellman}),
the optimal stopping rule sets thresholds on the Bayes factor
\citep{arrow1949}. For cost functions whose terminal payoffs are affine in the posterior, a BF threshold rule is optimal (\ref{app:bf_threshold}).

This covers misclassification costs, customer-impact utilities
\citep{wan2023}, and other sequential formulations where the payoff
scales with the expected treatment effect. In a related but
non-sequential framework, \citet{feit2019} optimize a fixed test
sample size to maximize expected profit.
The most common practical setup is
a directional stopping rule with a symmetric continuous prior
$\delta \sim N(0, V^2)$: stop on the Bayes factor, then check
the sign of the posterior mean. Here the affine structure only holds approximately, and the rule is near-optimal rather than exactly optimal (\ref{app:bf_threshold}). Specifically, it is exactly optimal under the two-point model when false-positive and false-negative costs are equal, and numerical
evidence suggests only a small gap under moderate cost asymmetry.

The expected opportunity loss rule of \citet{stucchio2015} is a
decision-theoretic stopping rule whose loss function is not affine in
the posterior. Its stopping criterion $E[\max(\delta, 0)]$ involves a
nonlinear function of the posterior (a truncated expectation), so the
BF threshold result does not apply and the resulting stopping
boundaries differ qualitatively from BF stopping.

\section{A Three-Tier Guarantee Hierarchy}
\label{sec:hierarchy}
Much of the recent work on Bayesian A/B testing has focused on
achieving frequentist error-rate properties while retaining Bayesian
advantages such as coherent belief updating, shrinkage, and evidence
for the null. To structure the overview, we organize popular
configurations into three tiers of statistical guarantees, ordered by
the strength of their error-rate control. Each higher tier adds
guarantees rather than replacing those below it. The tiers classify
stopping rules by their error-rate properties, not by overall quality.
For example, decision-theoretic methods whose stopping criteria are
posterior functionals (such as expected-loss stopping) provide no
error-rate guarantee and therefore appear in Tier~1, but when their
cost model is accurate they are optimal for the losses they target.
Table~\ref{tab:decision_matrix} maps the full space of meaningful
(prior, stopping rule) configurations against the resulting guarantees.

\begin{table*}[t]
\footnotesize
\setlength{\tabcolsep}{4pt}
\centering
\caption{Decision matrix for Bayesian A/B testing configurations. Each
row is a (prior, stopping rule) combination. BFDA = Bayes Factor Design
Analysis \citep{schonbrodt2018}. ``Partial'' (FPR correction): the EB
floor reduces but does not eliminate the $K$-metric inflation
(Section~\ref{sec:tier3}).}
\label{tab:decision_matrix}
\begin{tabular}{p{1.5cm} p{2.0cm} p{1.4cm} p{1.5cm} p{1.5cm} p{1.4cm} p{2.0cm} p{3.3cm}}
\toprule
Prior & Stopping rule & FPR ${\leq}\alpha$ & FPR corr.\ needed & FDR corr.\ needed & Shrinkage & Power bounded & Notes \\
\midrule
\multicolumn{8}{l}{\textit{Fixed-$n$ baseline (no optional stopping)}} \\[2pt]
Flat & Fixed-$n$ & Yes ($=\alpha$) & Yes & Yes & None & Yes (closed form) & Numerically identical to the z-test \\[3pt]
\midrule
\multicolumn{8}{l}{\textit{\textbf{Tier 1}: Posterior coherence (no frequentist error-rate guarantees)}} \\[2pt]
Flat & $P(B{>}A)>\gamma$ & No (inflated) & Yes & Yes & None & No & Identical to naive peeking \\[3pt]
Flat & $E[\max(\delta,0)]<\varepsilon$ & No (FPR$\approx$0.50) & Yes & Yes & None & No & Decision-theoretic, $\varepsilon$ controls precision, not FPR \\[3pt]
JZS/Cauchy & $P(B{>}A)>\gamma$ & No (inflated) & Yes & Yes & Weak & No & Valid posterior, FPR uncontrolled \\[3pt]
EB mixture & $P(B{>}A)>\gamma$ & No (inflated) & Yes & Yes & Calibrated & No & Shrinkage without FPR guarantee \\[3pt]
EB mixture & $E[\max(\delta,0)]<\varepsilon$ & No (uncontrolled) & Yes & Yes & Calibrated & No & Same, FDR requires BF stopping \\[3pt]
\midrule
\multicolumn{8}{l}{\textit{\textbf{Tier 2}: Bounded FPR via Bayes factor stopping}} \\[2pt]
JZS/Cauchy & BF $>1/\alpha$ & Yes & Yes & Yes & Weak & Yes (BFDA sim.) & No corpus, one parameter ($V=\delta_{\mathrm{MDE}}/\sigma$) \\[3pt]
Informative & BF $>1/\alpha$ & Yes & Yes & Yes & Toward $\mu_0$ & Yes (BFDA sim.) & Power and posterior sensitive to prior accuracy \\[3pt]
\midrule
\multicolumn{8}{l}{\textit{\textbf{Tier 3}: Per-metric FDR calibration and calibrated shrinkage via empirical Bayes (corpus ${\geq}200$ per program required)}} \\[2pt]
EB mixture & BF $>1/\alpha$ & Per-metric (Ville) & Partial & Per-metric & Calibrated & Yes (BFDA sim.) & Metric-level FDR via posterior odds \\
\bottomrule
\end{tabular}
\end{table*}

\subsection{Tier 1: Posterior coherence}\label{sec:tier1}
Any procedure that produces a proper posterior achieves this tier. The
Likelihood Principle \citep{bw1988} implies that the posterior is a
valid belief update regardless of the stopping rule used to collect the
data. Whether the analyst stopped after a fixed sample, after a
continuous monitoring rule triggered, or opportunistically on a weekend,
the posterior $\pi(\delta \mid \text{data})$ is coherently defined for
the observed data.

Posterior coherence does not provide any frequentist error-rate
guarantee on a stopping decision based on that posterior. Under a flat
prior, stopping the first time $P(B > A) > \gamma$ with $\gamma = 0.95$
exactly mirrors naive frequentist peeking
(Section~\ref{sub:flat_equiv}). Informative priors shift the realized
FPR but do not bound it. A common convention is to set $\gamma = 1 - \alpha$. This creates a
persistent source of confusion, because the $\alpha$ in the posterior
threshold $P(B > A) > 1 - \alpha$ is a Bayesian decision parameter
with no frequentist guarantee attached. It is not the same quantity as
the $\alpha$ in the frequentist bound $\mathrm{FPR} \leq \alpha$,
even when they take the same numerical value. The two coincide only
under a flat prior in a fixed-sample design
(Section~\ref{sub:flat_equiv}). Under optional stopping they diverge,
and the shared notation obscures this.

Any stopping rule whose criterion is a direct function of the
posterior is Tier~1 by construction. This includes two very different
kinds of configurations. Posterior-probability stopping provides
neither error-rate control nor a clear decision-theoretic justification
for optional stopping. Expected-loss stopping is different. It
explicitly targets a well-defined loss, and when the cost model
describes the true costs well it is optimal for that loss. The
hierarchy does not measure the dimension it optimizes for. Criteria in
the higher tiers are all based on the same function of the posterior,
the Bayes factor.

\subsection{Tier 2: Bounded FPR via Bayes factor stopping}
\label{sec:tier2}

This tier requires $\mathrm{FPR} \leq \alpha$ unconditionally over
repeated experiments, including under optional stopping. It is achieved
by combining any proper prior with Bayes factor thresholding as the
stopping rule. BF stopping requires a proper prior
(Section~\ref{sub:bf_stopping}). Under a flat prior the BF is undefined
and Tier~2 is out of reach.

\subsubsection{Type~I control via Ville's inequality}

For any proper prior $\pi$, the BF sequence
$\{\mathrm{BF}_{10,n}\}$ is a nonneg\-ative martingale under $H_0$
\citep{ville1939,rouder2014}. Ville's inequality then gives
\begin{equation}
  P\!\left(\sup_n \mathrm{BF}_{10,n} \geq 1/\alpha \;\middle|\; H_0\right) \leq \alpha,
\end{equation}
which is the anytime-valid form. The BF can be checked after every
observation and the bound holds at any stopping time without inflating
the Type~I rate. No $n_{\max}$ need be committed to in advance.

\subsubsection{Decision-theoretic optimality}
\label{sec:dt_optimality}

As noted in Section~\ref{sub:decision_theoretic}, it is often possible to approximate the optimal stopping rule for cost functions with affine terminal payoffs
(\ref{app:bf_threshold}) with BF thresholds. The BF is a likelihood ratio, and the same
likelihood-ratio structure that makes it a good decision statistic
also makes it a martingale under the null, which is why Ville's
inequality gives FPR control. Thresholding the BF on both sides
extends this to simultaneous control of both error rates: the upper
threshold $1/\alpha$ bounds FPR, and the lower threshold $\beta$
bounds prior-averaged FNR (\ref{app:fnr_bound}). Unlike the FPR
bound, which is unconditional, the FNR bound is averaged over the
prior distribution of effect sizes and applies only to paths that
cross the lower threshold. Paths reaching $n_{\max}$ without crossing
either boundary are additional false negatives not captured by the
bound.

\subsubsection{Evidence for the null}
\label{sec:evidence_null}

The lower threshold gives a principled way to abandon experiments
that are unlikely to show an effect, using the same BF that controls
the rejection decision. The FNR bound can be loose for small effects
near zero.

\subsubsection{Choosing a prior for BF stopping}
\label{sub:prior_choice}

For a program without a historical corpus, there are three common options for BF stopping.
The first is a weakly informative Jeffreys--Zellner--Siow (JZS) Cauchy prior \citep{rouder2009} with scale
$V = \delta_{\mathrm{MDE}}/\sigma$ per metric class. It requires no
historical data and reaches Tier~2. The second is a subjective
informative prior $N(\mu_0, s^2)$ that encodes domain expertise. Because
effect sizes in online experimentation are hard to predict in advance,
prior elicitation requires care; \citet{gronau2021} describes a quantile-matching procedure
tailored to the A/B test setting, and for a broader treatment see \citealt{garthwaite2005}. The third, available when sibling programs in the same organization already have calibrated
EB priors, is to borrow their $(\hat{p}, \hat{V})$ as a starting
prior. This treats the platform as a hierarchical model, but provides
Tier~3 guarantees only once a representative corpus has accumulated
for the new program.

\subsection{Tier 3: Bounded FDR and calibrated shrinkage via empirical Bayes}
\label{sec:tier3}

At Tier~2 there are three issues: multiplicity across the $K$ metrics in each experiment needs explicit correction (e.g., Bonferroni), the shrinkage
depends on whatever prior the analyst chose, and the false discovery
rate across experiments is uncontrolled. Tier~3 addresses all three. The empirical Bayes mixture prior partially absorbs the multiplicity
correction, calibrates the shrinkage to the program's actual
effect-size distribution, and controls the FDR. All Tier~2 guarantees are preserved. These additional properties depend on how well the EB prior represents
the effect-size distribution of the metrics in the experiments using it.

Within the BF stopping framework, the only way to reach Tier 3 is to add a calibrated empirical Bayes (EB) prior, estimated from a corpus of past A/B
  tests in the same program. \citet{deng2015} introduced an EB prior for A/B testing, later refined by \citet{deng2018}, in the form of a two-component normal mixture fitted to a corpus of
historical metric outcomes from past A/B tests in the program. The mixture
has a point mass at zero, representing the fraction $1-p$ of historical
metric observations that are null, plus $N(0, V^2)$ for non-null effects.
Many platforms use a simpler $N(0, V^2)$ prior without the point-mass
component. This still provides Tier~2 error control and shrinkage, but
the shrinkage is uncalibrated (the amount depends on the analyst's
choice of $V$ rather than the program's actual effect distribution).
The pooling assumption treats every metric in the program as an
exchangeable draw from this single mixture. It breaks down when metrics
have systematically different effect-size distributions
(Section~\ref{sub:corpus}).

\subsubsection{FDR calibration}

The BF threshold alone cannot control the FDR because it has no knowledge of
the non-null rate $p$. Under a ship-if-significant rule, the expected fraction
of shipped features with no true effect grows with the number of tests and
depends on $p$ directly. Formally,
\begin{equation}
\label{eq:fdr_approx}
  \mathrm{FDR} = \mathbb{E}\!\left[\frac{F}{\max(R,1)}\right]
  \approx \frac{(1-p)\,\alpha}{(1-p)\,\alpha + p\,(1-\beta)},
\end{equation}
where $F$ is false discoveries, $R$ is total discoveries, $1-\beta$ is marginal
power, and $p$ is the non-null rate as defined in
Section~\ref{sec:tier3} \citep{storey2003}. Given $\alpha$, $p$, and
power, the expected FDR is determined. At $p = 0.30$,
$\alpha = 0.05$, and power $\approx 0.85$, this gives
$\mathrm{FDR} \approx 0.12$.

The empirical Bayes prior addresses this by encoding the historical
metric-level non-null rate directly, for example by pooling across all
metrics in the program. A single $(\hat{p}, \hat{V})$ is fitted on the
corpus and applied identically to each of the $K$ metrics in new experiments within the program.

With the prior in place, BF stopping on the mixture e-process
$\Lambda_{n,k} = (1-\hat{p}) + \hat{p}\cdot\mathrm{BF}_{n,k}$
delivers per-metric FDR calibration without an explicit Bonferroni or
Benjamini-Hochberg correction. The mechanism is a posterior-odds
argument: the $(1-\hat{p})$ floor ensures that a rejected metric has
posterior null probability at most $(1-\hat{p})\,\alpha$, bounding the
Bayesian FDR \citep{storey2003,efron2008,scott2010}. Because the
mixture inherits the martingale property from the BF, this bound holds
at any stopping time. Under correct model specification, the Bayesian
and frequentist FDR coincide \citep{storey2003}.
Section~\ref{sec:sim} confirms experiment-level FDR behavior
under various corpus conditions.

\textbf{Implications for FPR.}
The $(1-\hat{p})$ floor also tightens per-metric FPR as a side effect.
Most platforms ship when any metric crosses the threshold. Under this
ship-if-any rule, the experiment-level FPR grows with $K$, and the
floor substantially reduces it. The reduction depends on how well
$\hat{p}$ represents the true non-null rate. Empirical examples are in
Section~\ref{sec:sim}.

\subsubsection{Calibrated shrinkage}
As noted in Section~\ref{sub:posterior_stopping}, any informative prior
produces shrinkage. What the EB prior adds is \emph{calibration}: the
shrinkage factor is fitted to the program's actual effect-size
distribution rather than chosen by the analyst. Under the EB mixture, the posterior mean is shrunk toward zero by a
factor that depends on the prior variance and the metric's observation
noise, with noisier metrics shrunk more \citep{deng2015}. This
partially corrects the Type~M (winner's curse) inflation in estimates
selected at a significance threshold
(\ref{app:typeM}; \citealt{kessler2024}).

Shrinkage also reduces the variance of the effect estimator.
This reduction is orthogonal to regression adjustment (CUPED), which
removes pre-experiment variance through covariate adjustment. The two
can be applied simultaneously \citep{deng2015,deng2023b}. Unlike
CUPED, which only improves power, the EB prior pulls in two directions:
the variance reduction raises power, while the $(1-\hat{p})$ floor
raises the stopping threshold and can lower
power (Table~\ref{tab:simB}). Details on shrinkage and the James--Stein
connection are in~\ref{app:eb}.

\subsubsection{Prior estimation and corpus quality}
\label{sub:corpus}

All Tier~3 properties depend on how well $(\hat{p}, \hat{V})$
represents the metrics in the experiment. The parameters are estimated
offline from a corpus of historical per-metric A/B test observations
from the same program
\citep{dempster1977,efron2008,deng2015}.

\begin{figure}[t]
  \centering
  \includegraphics[width=\linewidth]{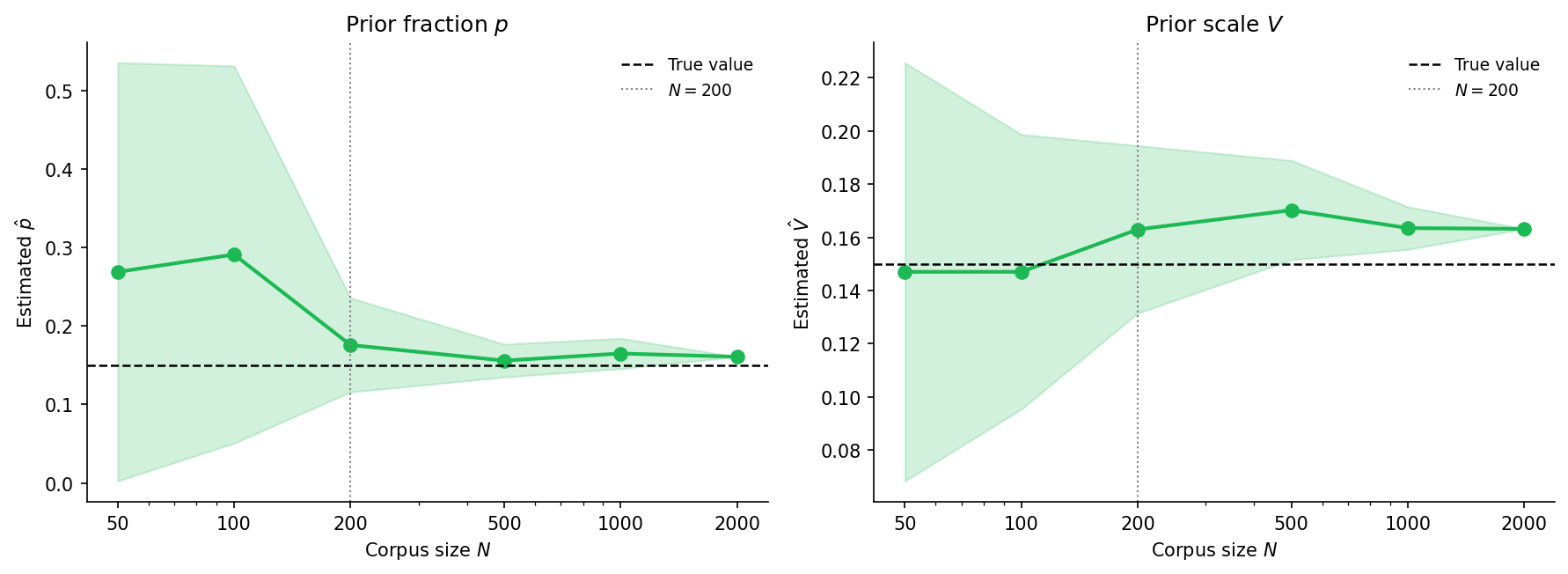}
  \caption{Estimation variance of EB prior parameters $p$ (left) and $V$
    (right) as a function of corpus size $K_c$. Both stabilize around
    $K_c = 200$ (dashed line). $\hat{p}$ is more sensitive to corpus
    size than $\hat{V}$.}
  \label{fig:prior_stability}
\end{figure}

The corpus needs to be large enough and representative. Calibration
reliability stabilizes around $K_c \approx 200$ historical experiments
per program \citep{deng2015} (Figure~\ref{fig:prior_stability}). Since
metric results within an experiment are correlated, many metrics per
experiment cannot substitute for few experiments.

When the corpus is not representative, the error-rate guarantees
degrade but the MSE advantage from shrinkage is more robust: even a
misspecified prior partially corrects Type~M inflation
(Table~\ref{tab:simB}). Three failure modes illustrate this.
\textbf{Winner-selected corpus}: excluding non-significant experiments
causes $\hat{p} \to 1$, collapsing the $(1-\hat{p})$ floor entirely.
\textbf{Pooled programs}: pooling programs with different non-null rates
produces $\hat{p}_{\mathrm{pooled}}$ equal to a weighted average of the
constituent non-null rates. A program with a high null rate inherits an
inflated $\hat{p}$ that weakens multiplicity correction, while a program
with a low null rate is over-corrected.
\textbf{Heterogeneous metrics}: pooling metrics with different
effect-size distributions over-shrinks easy-to-move metrics and
under-shrinks hard ones \citep{stephens2017,dimmery2019}. None of
these can be fixed by collecting more data of the same kind. All three
are studied in the simulation (Section~\ref{sec:sim}).

\subsubsection{Robustness to prior misspecification}
\label{sub:robustness}

When the EB prior may be misspecified, the Bayesian robustness
literature offers several approaches to handling prior misspecification.
\citet{berger1986} embed the EB prior in an $\varepsilon$-contamination
class and compute posterior bounds over that class, with a
user-specified cap on how far the posterior can deviate from the base
prior. \citet{egidi2022} mix the informative prior with a diffuse
fallback and learn the mixing weight from the data, discarding the
informative prior entirely when conflict is detected.
\citet{grunwald2024} show that e-values provide structural robustness
of the Type~I guarantee to prior choice, since the martingale property
holds for any proper prior. All three preserve Tier~2 FPR control (Ville
holds for any proper prior). Shrinkage degrades smoothly. FDR
calibration, which depends on the $(1-\hat{p})$ floor being at the
right level, is harder to recover and depends on the specifics of the
robust mechanism. The details of these methods and their properties in the sequential
A/B testing setting are beyond the scope of this paper.

Programs below the corpus threshold should not use an underfitted EB
prior. Without a calibrated prior, a program needs explicit
multiple-testing corrections (e.g., Bonferroni across metrics and
adaptive BH across experiments; \citealt{storey2002,hartog2023}) and
should operate at Tier~2 until corpus integrity can be maintained.

\section{Practical Considerations}
\label{sec:complications}

\subsection{Sample size planning under optional stopping}
\label{sub:ss}

Power planning matters under any framework. Without a maximum sample size
$n_{\max}$, an underpowered experiment may run indefinitely at a cost that
cannot be bounded in advance. If $n_{\max}$ is set too low, the only
experiments that stop are those where noise happened to produce a large BF,
exactly the regime where effect estimates are most inflated. Power analysis
is necessary to bound both the detection rate and the estimation error
conditional on stopping.

The standard approach is Bayes Factor Design Analysis (BFDA;
\citealt{schonbrodt2018}), which simulates datasets under the prior
predictive and estimates power as the proportion that cross $1/\alpha$
before $n_{\max}$. For the normal-conjugate model, the power function
has a closed form \citep{johari2022,pawel2025,hagar2025}. For
non-conjugate models, BFDA requires simulation
\citep{stefan2019}.

Just like the inference from the experiment, the power analysis is
affected by how well the prior reflects the true effect distribution.
Prior sensitivity analysis over a plausible range of effect sizes is
therefore part of a responsible BFDA.

\subsection{Intervals after Bayes factor stopping}
\label{sub:ci}

Because BF stopping controls FPR in the frequentist sense, it is
possible to construct \emph{confidence} intervals that, as opposed to
credible intervals, are consistent with the stopping decision.
The confidence sequence is obtained by inverting the BF across
candidate null values \citep{wagenmakers2020}. Under the
normal-conjugate model this produces a closed-form interval
$\hat\delta_n \pm w_n$. For non-conjugate models the inversion
requires numerical computation. The interval excludes zero if and only
if the BF exceeds $1/\alpha$ \citep{grunwald2024}, so the interval
and the stopping decision always agree.

\begin{figure}[t]
  \centering
  \includegraphics[width=\linewidth]{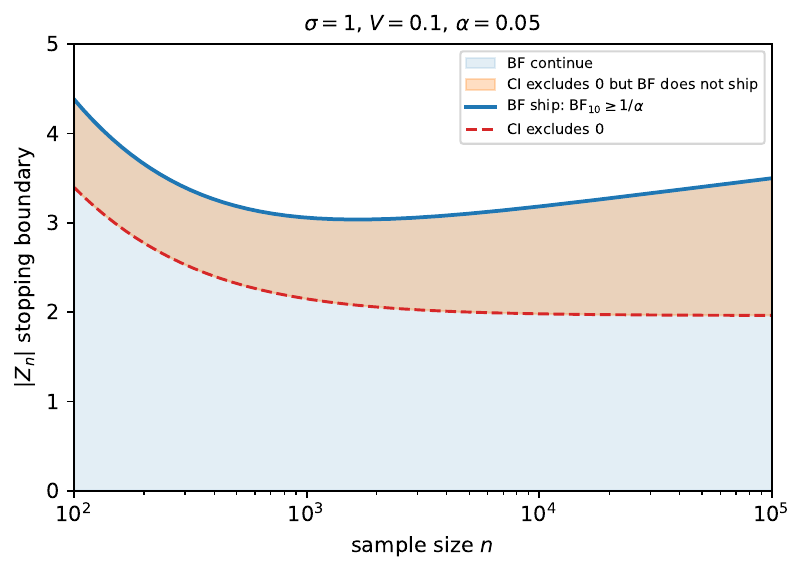}
  \caption{Stopping boundaries on $|Z_n|$ as a function of $n$ for the
    normal-conjugate model ($\sigma = 1$, $V = 0.1$, $\alpha = 0.05$).
    The blue region is the BF continuation zone. The orange band is
    where the credible interval excludes zero but the BF has not yet
    crossed $1/\alpha$.}
  \label{fig:bf_ci_boundaries}
\end{figure}

The credible interval and the confidence sequence answer different
questions. The credible interval is a probabilistic statement about
$\delta$ under the model. The confidence sequence carries a
frequentist coverage guarantee regardless of the stopping rule. In
the normal-conjugate model, when the BF crosses $1/\alpha$ the
credible interval also excludes zero
(\ref{app:bf_ci}; \citealt{campbell2024}), so both intervals are
consistent with the stopping decision.
Figure~\ref{fig:bf_ci_boundaries} illustrates this.

\section{Bayesian and Frequentist Connections}
\label{sec:freq_bayes_comparison}

For the configurations most commonly deployed in practice, Bayesian and
frequentist A/B testing procedures are more similar than the
long-running debate suggests. In many situations, the choice of configuration is likely to matter
more than the choice of framework.

\subsection{Flat-prior inference}
\label{sub:flat_equiv}

Under a flat prior and the two-group normal model, Bayesian and
frequentist procedures produce numerically identical outputs. The
posterior mean equals the MLE. The one-sided posterior probability
$P(\delta > 0 \mid \text{data})$ equals $1 - p_{\mathrm{val}}$. The
equal-tailed credible interval has the same endpoints as the
fixed-sample normal confidence interval. The threshold
$P(B > A) > 0.95$ is algebraically equivalent to the one-sided
rejection region $Z_n > z_{0.05} = 1.645$. These equalities hold
exactly for known $\sigma^2$ and asymptotically under standard
regularity conditions \citep{vandervaart1998}.

Under an informative prior centered at zero, the same fixed-$n$
posterior-probability threshold produces FPR below $\alpha$, because the
prior shrinks the posterior mean toward zero and raises the effective threshold
for $P(B > A) > \gamma$.

Several commercial experimentation platforms offer flat or near-flat
prior Bayesian inference as a default configuration
\citep{statsig2024,eppo2024,growthbook2024}. A platform deploying this configuration
with posterior-probability thresholds is running frequentist-equivalent
inference under a different vocabulary.

Flat-prior optional stopping is also identical to naive frequentist
peeking: ``stop when $P(B > A) > 0.95$'' is the same rule as ``stop
when $p_n < 0.05$.'' Our simulation confirms FPR $= 0.303$ in both
cases (Table~\ref{tab:simA}).

\subsection{Bayes factor stopping and the mSPRT}
\label{sub:msprt_equiv}

One of the most common sequential frequentist procedures is the mixture
Sequential Probability Ratio Test (mSPRT;
\citealt{robbins1970,johari2022}). This statistic is exactly the Bayes
factor under the same prior \citep{johari2022}. A frequentist who
selects a $N(0, V^2)$ mixing distribution to maximize power at the MDE
is, in Bayesian terms, placing a normal prior on the treatment effect.
The mixing distribution \emph{is} a prior. The choice affects power but
not validity, since the martingale property holds for any proper prior.

What distinguishes the mSPRT from BF stopping is point estimation. A
frequentist typically reports the MLE $\hat\delta$ at stopping. A
Bayesian reports the posterior mean, which shrinks toward zero. In the conjugate normal case, a frequentist could apply a ridge or
James--Stein estimator after stopping and obtain shrinkage of the same
order, so the distinction is one of default practice, not of what the
frameworks permit.

\subsection{Costs and error rates}
\label{sub:cost_rate}

Any deterministic, non-adaptive stopping rule applied independently
across experiments in a stationary program will converge to some FPR
and FNR. For many of the cost functions discussed in the A/B testing
literature, the relationship is tighter:
the optimal or near-optimal policy is a BF threshold rule
(Section~\ref{sub:decision_theoretic}, \ref{app:bf_threshold}), and
Ville's inequality directly converts the threshold into a prior-free
FPR bound. When this convergence holds, a frequentist who selects
$\alpha$ and $\beta$ by reasoning about costs has implicitly chosen
cost ratios, and, conversely, a Bayesian who specifies costs inherits error rates.

To illustrate, the utility of \citet{wan2023} accounts for customer
impact during the experiment, revenue over a post-launch planning
horizon, and a hurdle cost for launching. This utility thinks in
business terms, not statistical ones, yet BF stopping is often
near-optimal for it (\ref{app:bf_threshold}). For this class of cost functions,
the experimenter who specified business costs, not error rates, has
nonetheless chosen a point in $(\alpha, \beta)$ space.

The convergence does not hold universally. Expected-loss stopping
\citep{stucchio2015} is an illustrative counterexample.
Its criterion $E[\max(-\delta,0)] < \varepsilon$ is a precision
condition that triggers when the posterior is tight enough that the
expected harm is small. Under $H_0$, precision increases without
evidence for either hypothesis. The distinction is visible in the
stopping boundaries: the BF boundary in posterior-mean space shrinks
as $O(\sqrt{\log n / n})$, requiring progressively larger effects to
reject, while the expected-loss boundary converges to zero,
eventually triggering at any $\hat\delta > 0$ once the posterior
concentrates. Expected-loss stopping therefore ships whichever variant
is ahead once the posterior is sufficiently concentrated. Under a flat
prior, the resulting directional FPR is approximately $0.50$
(Table~\ref{tab:simA}). Any rule that always ships a variant and
selects whichever is ahead shares this FPR by symmetry of the null,
including the $P(B > A) > 0.50$ rule in Table~\ref{tab:simA}. The FPR
is the same, but the stopping times, power, and estimation accuracy
differ because the rules use different criteria to decide when to stop.
Expected-loss stopping gains detection
speed by spending the error budget that BF stopping reserves for
false-positive control. This trade-off is favorable when the operating characteristic is expected
regret per decision and the cost of a false positive is bounded by
$\varepsilon$ (as argued by proponents of this approach). When shipping incurs any non-metric cost, the regret ranking reverses
(Setting~D, Figure~\ref{fig:shipping_cost}).

\subsection{Implicit priors in frequentist estimation}
\label{sub:implicit_priors}

The convergence between frameworks extends to estimation. Frequentist
procedures routinely incorporate structural prior beliefs without
naming them as such. The James--Stein estimator \citep{james1961}
shrinks toward zero on the implicit assumption that effects are small,
encoding the same belief as a Gaussian prior. Regularized regression
adjustment \citep{bloniarz2016,wager2016} imposes beliefs about
covariate magnitudes through its penalty \citep{efron2016}. The
distinction between the frameworks is therefore not whether prior
information is used, but whether it is stated explicitly, as part of
a generative model whose posterior has a direct probabilistic
interpretation.

Many practitioners are aware of the relationship between flat-prior
Bayesian and frequentist methods. In this section we have shown that
the connections run deeper: the mSPRT and the BF are the same
statistic, cost-motivated and error-rate-motivated thresholds lead to
similar stopping rules, and regularized frequentist estimators can
produce the same shrinkage as Bayesian posteriors.

\section{Simulation Evidence}
\label{sec:sim}

Four simulations address different questions. Settings~A--C correspond
to the tier structure. Setting~A verifies the Tier~1/Tier~2 boundary:
which configurations control FPR and which do not. Setting~B evaluates
Tier~2 and Tier~3 configurations against frequentist sequential
alternatives on power, MSE, and stopping time. Setting~C tests how
Tier~3 properties depend on corpus quality. Setting~D evaluates
methods on the decision-theoretic metric that expected-loss stopping
optimizes for, showing how the ranking depends on the assumed cost of
shipping a null-effect variant. Full design parameters are in
\ref{app:sim}.

\subsection{Which Bayesian configurations control FPR (Setting~A)}
\label{sub:sim:comparison}

Table~\ref{tab:simA} illustrates the baseline ordering by crossing stopping
rules against prior families on a normal DGP ($\delta = 0.2\sigma$, continuous
monitoring, $n_{\max} = 5{,}000$ per arm). The purpose is to separate
configurations that control FPR from those that do not. The relevant
comparators are the valid sequential frequentist methods treated in Setting~B,
not a fixed-$n$ test.

\begin{table*}[t]
\centering
\caption{Setting~A: FPR control across stopping rules and priors. Normal DGP
  ($\sigma=1$, $\delta=0.2$ under $H_1$, design details in
  \ref{app:sim}).
  \emph{FPR}: $P(\text{declare }B\text{ wins}\mid H_0)$, directional.
  \emph{Pwr@}$n_{\max}$: $P(\text{stop before }n{=}5{,}000\mid H_1)$.
  \emph{Avg.}~$n$: conditional mean stopping time among $H_1$ experiments
  that stopped. Naive peeking is included as a calibration check only.}
\label{tab:simA}
\begin{tabular}{llrrr}
\toprule
Prior & Stopping rule & FPR & Pwr@$n_{\max}$ & Avg.\ $n$ \\
\midrule
\multicolumn{2}{l}{\textit{Calibration baseline (not a valid sequential procedure)}} \\
None & Naive peeking ($Z_n>1.645$)            & 0.303 & 1.000 & 185 \\
\midrule
\multicolumn{2}{l}{\textit{Flat prior ($\pi(\delta)\propto 1$)}} \\
Flat & $P(B{>}A)>0.95$                       & 0.303 & 1.000 & 185 \\
Flat & $P(B{>}A)>0.50$ ($n_{\min}{=}350$)      & 0.501 & 0.996 & 350 \\
Flat & Expected loss ($\varepsilon=0.02$)    & 0.499 & 0.954 &  85 \\
\midrule
\multicolumn{2}{l}{\textit{BF stopping ($\Lambda_n>1/\alpha$), varying prior}} \\
Normal ($V=\delta_{\mathrm{MDE}}/\sigma$)         & BF & 0.019 & 1.000 & 502 \\
Weak Gaussian ($V=2\delta_{\mathrm{MDE}}/\sigma$)  & BF & 0.016 & 1.000 & 525 \\
\bottomrule
\end{tabular}
\end{table*}

Flat-prior $P(B > A) > 0.95$ and naive peeking are algebraically identical
(FPR $= 0.303$), confirming the equivalence from Section~\ref{sub:flat_equiv}.
Expected-loss stopping \citep{stucchio2015}, which was never intended to
bound the FPR, is included to illustrate Tier~1 configurations and the
observation that any stopping policy has \emph{some} FPR across the
program. Here that rate is approximately $0.50$.
The $P(B > A) > 0.50$ rule with a minimum sampling period achieves
FPR $= 0.501$, consistent with the precision-criterion interpretation
from Section~\ref{sub:cost_rate}.
BF stopping controls FPR across all proper prior choices
(FPR $= 0.016$--$0.019$). These rates are well below $\alpha = 0.05$
partly because the BF is a two-sided statistic while the decision is
directional (declare $B$ wins), which halves the FPR. The point
null $\delta = 0$ is the least favorable case in the one-sided null
$\delta \leq 0$, so the Ville bound applies to the directional
decision.

\begin{figure}[t]
  \centering
  \includegraphics[width=\linewidth]{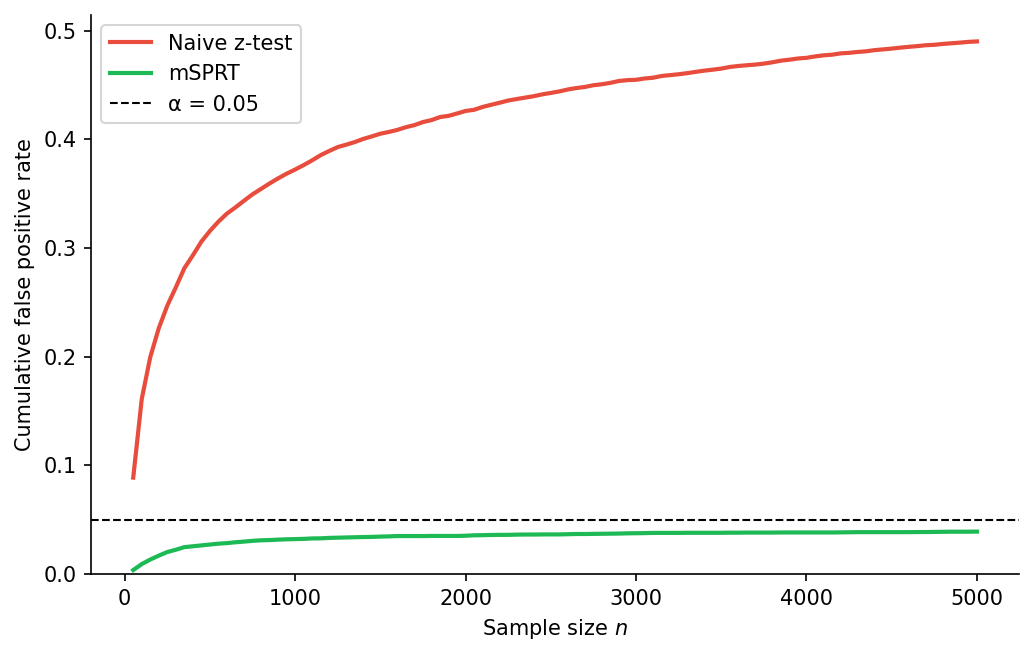}
  \caption{Cumulative FPR under continuous monitoring ($H_0$ true).
    The naive z-test FPR rises monotonically, while the mSPRT (equivalent
    to Gaussian-conjugate BF stopping) FPR stays below $\alpha=0.05$.}
  \label{fig:msprt}
\end{figure}

Figure~\ref{fig:msprt} shows cumulative FPR trajectories under continuous
monitoring. The naive z-test rises monotonically. BF stopping stays below
$\alpha = 0.05$ throughout, consistent with the supermartingale property.

\begin{figure}[t]
  \centering
  \includegraphics[width=\linewidth]{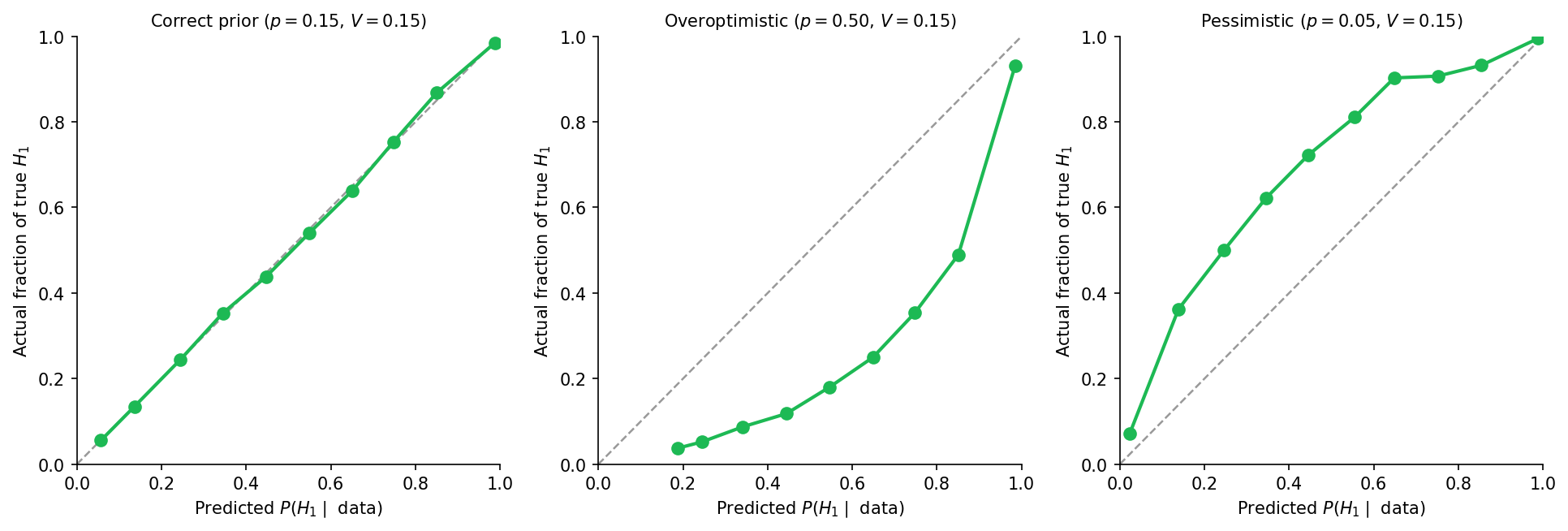}
  \caption{Reliability diagrams for three prior specifications under BF
    stopping. Well-calibrated (left), overoptimistic ($p=0.50$, center),
    and pessimistic ($p=0.05$, right). Prior miscalibration corrupts posterior
    calibration without affecting FPR control.}
  \label{fig:sensitivity}
\end{figure}

Figure~\ref{fig:sensitivity} shows reliability diagrams for three prior
specifications under BF stopping. Prior miscalibration corrupts posterior
reliability without affecting the Type~I guarantee from BF stopping.

\subsection{BF stopping vs.\ sequential frequentist methods (Setting~B)}
\label{sub:sim:sequential}
To contrast the Bayesian optional stopping configurations, Setting~B asks how BF
stopping compares to the frequentist sequential procedures that mature platforms
actually deploy as defaults, namely O'Brien--Fleming GST \citep{obrien1979} and
always-valid frequentist inference via Waudby-Smith--Ramdas confidence sequences
\citep{waudby2024}. All three methods provide unconditional Type~I control. The
simulation evaluates FPR, power, average stopping time, and point-estimate MSE
across seven prior and method conditions on a common DGP ($\sigma = 1$, $\delta
= 0.2$ under $H_1$, $n_{\max} = 1{,}000$ per arm, $p = 0.30$,
$n_{\mathrm{sims}} = 5{,}000$).

GST requires committing to $n_{\max}$ in advance (Lan--DeMets
$\alpha$-spending allows flexible look times within that horizon). We
report two realistic schedules, 14 equally spaced looks (daily
monitoring over a 2-week experiment) and 21 looks (3-week experiment),
with boundaries computed by one-sided OBF $\alpha$-spending. WS--R and
BF stopping are anytime-valid and require no $n_{\max}$ commitment. We give the always-valid methods their best
case, monitoring after every observation pair, to make the comparison fair
to them. For BF stopping we test four conditions. First, a Gaussian prior
$N(0, V^2)$ with $V = \delta_{\mathrm{MDE}}/\sigma$, representing a Tier~2 configuration
with no historical data. Second, an oracle EB prior using the true
program parameters ($p = 0.30$, $V = 0.20$), representing the best
possible Tier~3 configuration. Third, an EB prior estimated from a
winner-selected corpus where only significant results were retained,
causing $\hat{p} \to 1$ and the $(1-\hat{p})$ floor to collapse.
Fourth, an EB prior from a corpus that pools two programs with
different effect-size distributions ($\hat{p} \approx 0.51$).

\begin{table*}[t]
\footnotesize
\setlength{\tabcolsep}{4pt}
\centering
\caption{Setting~B: sequential methods comparison. Common DGP,
  $n_{\max}=1{,}000$ per arm. FPR is experiment-level and directional.
  Pwr@$n_{\max}$ is $P(\text{stop before }n_{\max} \mid H_1)$. Avg.~$n$ is
  conditional on stopping under $H_1$. MSE uses the posterior mean for
  Bayesian methods and $\hat\delta$ for frequentist. Int.\ width is the
  mean reported-interval width at stopping.}
\label{tab:simB}
\begin{tabular}{lp{2.8cm}rrrrrr}
\toprule
Method & Prior / config & FPR & Pwr@$n_{\max}$ & Avg.\ $n$ & MSE $\times 10^3$ & Int.\ width & $n_{\max}$ commit? \\
\midrule
\multicolumn{8}{l}{\textit{Frequentist sequential}} \\
GST & OBF, 14 looks (daily, 2 wk) & 0.051 & 0.996 & 430 & 5.9 & 0.395 & Yes \\
GST & OBF, 21 looks (daily, 3 wk) & 0.052 & 0.997 & 421 & 6.3 & 0.408 & Yes \\
Always-valid & WS--R CS ($\rho^2 = \delta_{\mathrm{MDE}}^2$) & 0.013 & 0.955 & 392 & 17.3 & 0.521 & No \\
\midrule
\multicolumn{8}{l}{\textit{BF stopping (Tier 2)}} \\
BF & Gaussian ($V=\delta_{\mathrm{MDE}}/\sigma$) & 0.011 & 0.941 & 405 & 3.0 & 0.297 & No \\
\midrule
\multicolumn{8}{l}{\textit{BF stopping (Tier 3, EB prior)}} \\
BF & EB oracle ($p=0.30$, $V=0.20$)   & 0.003 & 0.899 & 478 & \textbf{2.7} & 0.299 & No \\
BF & EB winner ($\hat{p}\to 1$)        & 0.013 & 0.953 & 388 & 4.5 & 0.306 & No \\
BF & EB pooled ($\hat{p}\approx 0.51$, $\hat{V}\approx 0.37$) & 0.007 & 0.910 & 432 & 10.4 & 0.337 & No \\
\bottomrule
\end{tabular}
\end{table*}

Three findings stand out in Table \ref{tab:simB}. First, all valid
sequential methods control FPR at or below the nominal $\alpha = 0.05$.
Among them, GST with realistic monitoring achieves the highest power,
including higher power than BF stopping with the oracle EB prior (the
best possible prior for this program). BF stopping with a Gaussian prior is
slightly less powerful but stops earlier on the cases it does detect,
and the WS--R CS (tuned to the MDE for a fair comparison) is competitive
on power and stopping time but has substantially higher MSE. The relationship behind
this ordering is the threshold structure. GST's spending function
approaches the fixed-test $z$-threshold at the late looks, so paths with
marginal evidence eventually cross. The BF $|Z_n|$-threshold increases slowly (logarithmically)
over time, making it progressively harder to reject at later looks,
so it catches strong-signal paths quickly but never crosses on
marginal-signal paths. BF therefore detects the easy cases fast and misses the hard ones,
while GST is the opposite. This characterization reflects OBF
spending. Less conservative spending functions would narrow the
stopping-time gap.

The meaningful operational advantage of BF stopping over GST is
flexibility. BF stopping requires no $n_{\max}$ commitment, whereas GST
requires fixing $n_{\max}$ in advance (Section~\ref{sub:ss}).
The WS--R CS shares this flexibility. With the mixing variance tuned
to the MDE ($\rho^2 = \delta_{\mathrm{MDE}}^2$), it achieves power
and stopping time comparable to BF stopping, but at substantially
higher MSE ($17.3$ vs $3.0$ for BF/Gaussian) because the confidence
sequence does not shrink the point estimate.

Second, the oracle EB prior at Tier~3 achieves the lowest MSE of any
configuration. The reduction comes from reporting the posterior mean
(which shrinks estimates toward zero) rather than the MLE. This is a
property of the estimator, not the stopping rule, and in principle a
frequentist could apply shrinkage after GST or WS--R stopping as well.
In practice, BF stopping and the EB prior naturally produce the
posterior mean, while frequentist methods default to the MLE. FPR remains controlled and power is
comparable to the Gaussian configuration. The MSE reduction at Tier~3 with a
well-calibrated prior is the primary operational advantage of EB over
any Tier~2 or frequentist sequential alternative.

Third, the misspecified EB priors preserve FPR control (the martingale
property holds for any proper prior) but lose the MSE advantage, and in
one case \emph{reverse} it. The winner-selected corpus ($\hat{p} \to 1$)
recovers an EB variance close to the oracle, so its MSE is similar to the
uncorrected frequentist methods rather than oracle-level. The
pooled-programs corpus over-estimates $V$, which weakens posterior
shrinkage and produces \emph{higher} MSE than even uncorrected GST. This
confirms that the Tier~3 estimation advantage is inseparable from corpus
quality.

\begin{figure}[t]
  \centering
  \includegraphics[width=\linewidth]{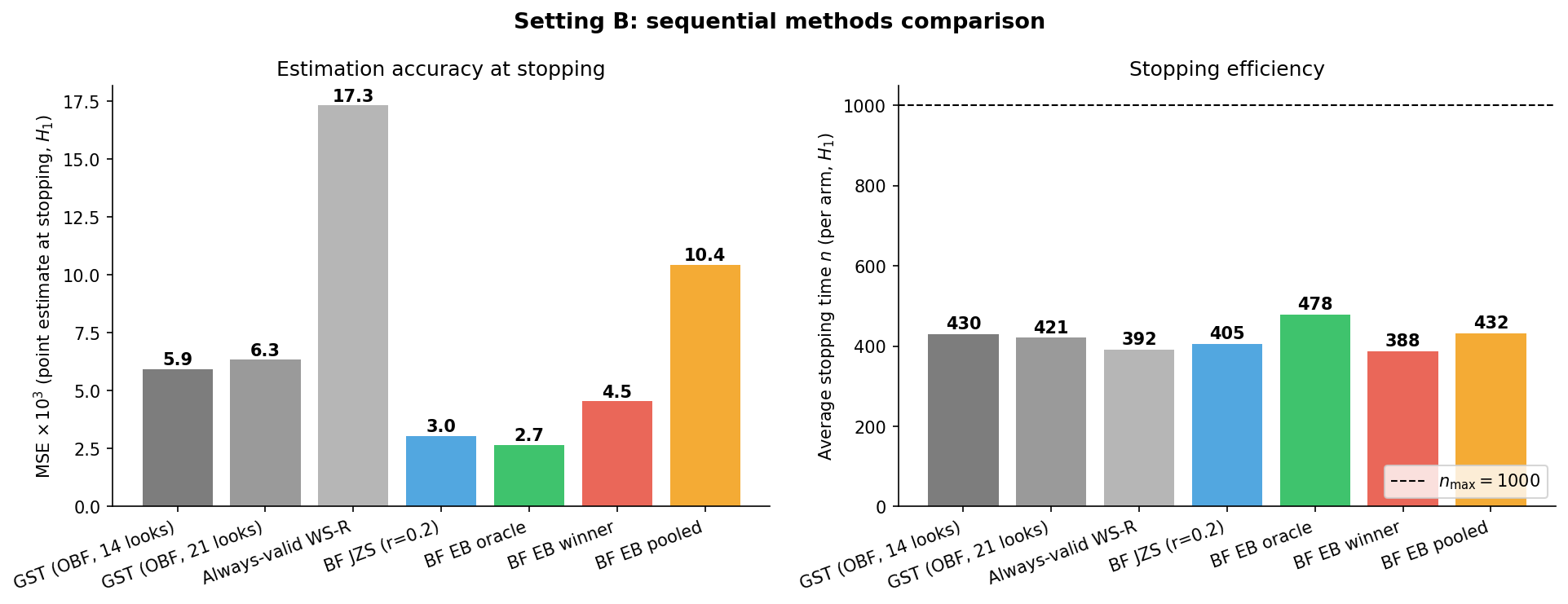}
  \caption{Setting~B: point-estimate MSE (left) and average stopping time
    (right) for each method and prior condition. The oracle EB prior achieves
    the lowest MSE. Misspecified EB priors keep FPR control but lose, or in
    one case reverse, the estimation advantage. All methods control FPR at or
    below $\alpha = 0.05$.}
  \label{fig:sequential_comparison}
\end{figure}

\subsection{EB prior: corpus size and quality (Setting~C)}
\label{sub:sim:eb}

Setting~C fits the EB mixture prior to a generated corpus and sweeps corpus size
$K_c \in \{10, 30, 50, 100, 500\}$ across four corpus quality conditions, namely
Oracle EB (true $p$, $V$ known), representative corpus, winner-selected corpus,
and pooled heterogeneous programs. The DGP is all-or-nothing at the experiment
level. Each experiment is fully null (all $K = 10$ metrics have $\delta = 0$)
with probability $1-p = 0.70$, or fully non-null ($\delta_k \sim N(0, V^2)$)
with probability $p = 0.30$. This simplification makes the experiment-level
non-null rate equal to the per-metric non-null rate $p$, so a single
corpus-derived $\hat{p}$ encodes both. Under a more realistic mixed DGP (each metric
independently null or non-null), the per-metric Ville bound is unchanged but the
experiment-level rates differ from the metric-level rates. We discuss the
implication in Section~\ref{sec:tier3}. Error rates are measured at the
experiment level.

\begin{figure}[t]
  \centering
  \includegraphics[width=\linewidth]{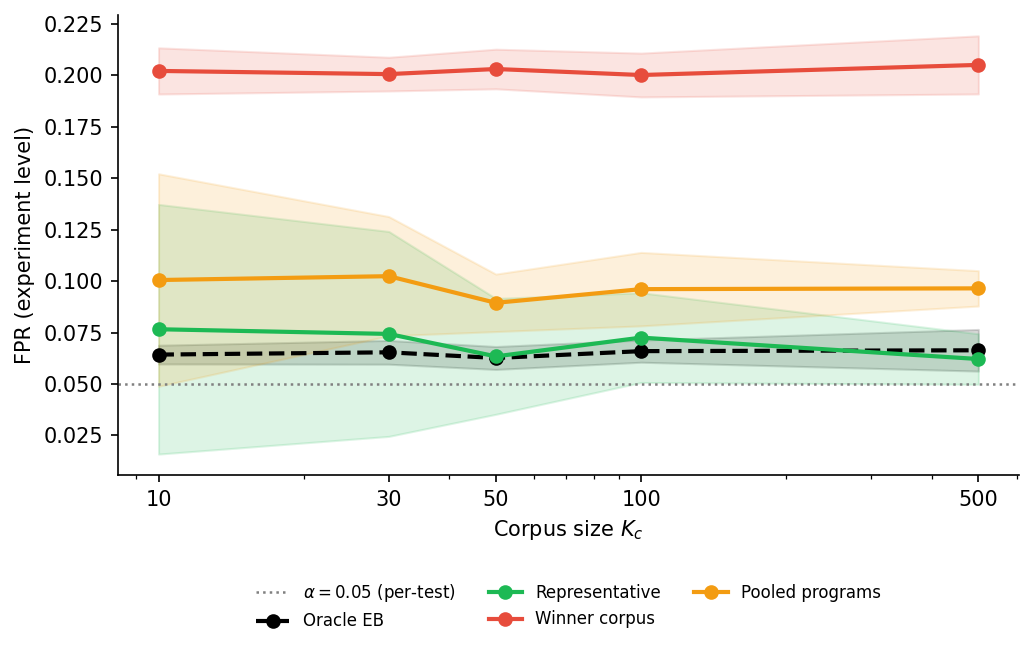}
  \caption{Setting~C, experiment-level FPR vs.\ corpus size $K_c$ ($K=10$
    metrics, $p=0.30$). Oracle EB sits at FPR $\approx 0.068$ for all
    $K_c$. The $(1-\hat{p})$ floor pulls the per-metric FPR down to
    $\approx 0.007$, and the experiment-level rate is
    $1-(1-0.007)^{10}\approx 0.068$ assuming independence across the
    $K$ metrics (correlated metrics would give a lower rate).
    Representative corpus converges to oracle by $K_c \approx 50$. Winner
    corpus stays at $\approx 0.20$ regardless of $K_c$, because the floor
    collapses when $\hat{p}\to 1$. Pooled programs stay elevated at
    $\approx 0.10$.}
  \label{fig:corpus_fpr}
\end{figure}

\begin{figure}[t]
  \centering
  \includegraphics[width=\linewidth]{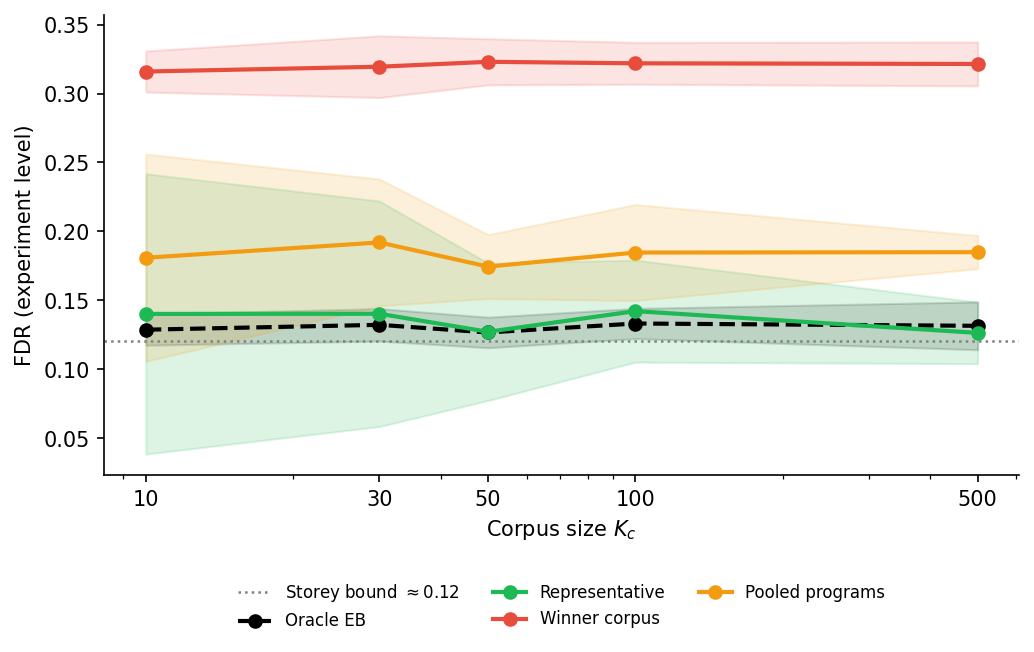}
  \caption{Setting~C: experiment-level FDR vs.\ corpus size $K_c$. The
    dotted line is the Storey approximation
    $\mathrm{FDR} \approx (1-p)\,\alpha / [(1-p)\,\alpha + p\,(1-\beta)]$
    (Eq.~\ref{eq:fdr_approx}) at $p = 0.30$, $\alpha = 0.05$, and an
    average per-test power of $0.85$, giving $\approx 0.12$. Oracle EB and
    representative corpora track this bound at $\approx 0.13$. Winner and
    pooled corpora exceed it at every $K_c$.}
  \label{fig:corpus_fdr}
\end{figure}

\begin{figure}[t]
  \centering
  \includegraphics[width=\linewidth]{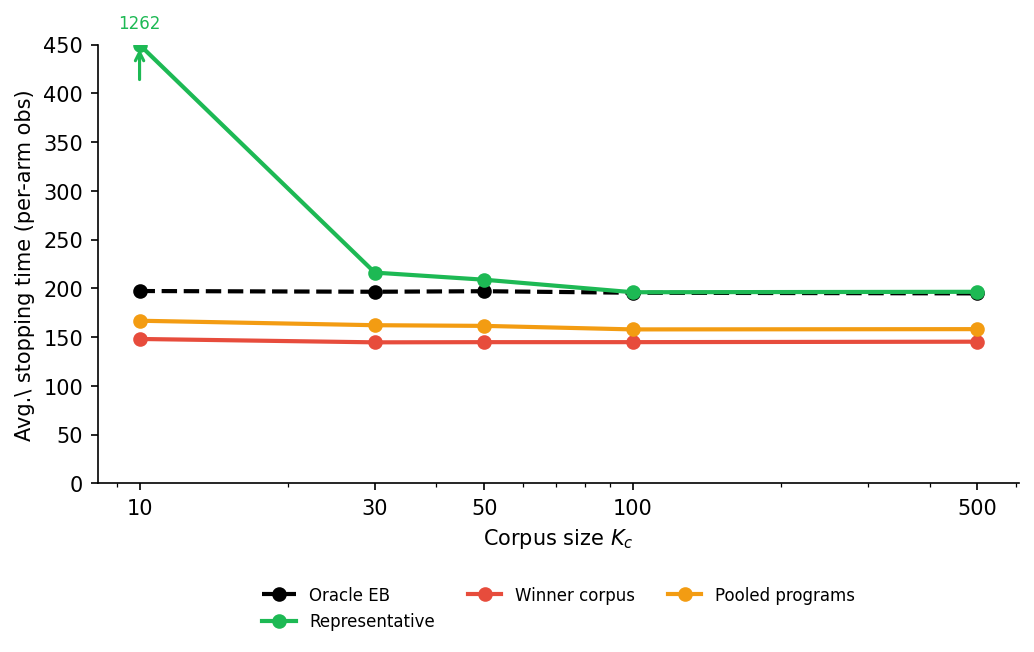}
  \caption{Setting~C, average stopping time vs.\ corpus size $K_c$
    (non-null shipped experiments). Oracle EB (threshold $\approx 64$)
    takes $\approx 195$ obs per arm. Winner corpus (threshold $\approx 20$,
    floor collapsed) stops $\approx 25\%$ faster, the same speed advantage
    that drives its inflated FPR. Pooled programs stop at $\approx 160$ obs.
    Representative at $K_c = 10$ occasionally reaches $\approx 1{,}260$ obs
    (arrow, truncated). A noisy $\hat{p}$ can produce a very high threshold.}
  \label{fig:corpus_avgn}
\end{figure}

The results confirm the failure-mode taxonomy from
Section~\ref{sub:corpus}. The relevant FDR target is not $\alpha$ but
the Storey approximation (Eq.~\ref{eq:fdr_approx}) evaluated at the
simulation parameters $p = 0.30$, $\alpha = 0.05$ and per-test power
$\approx 0.85$, which gives $\mathrm{FDR}\approx 0.12$ for any
Tier~3 procedure with a calibrated prior. Oracle EB and the
representative corpus track this line. The representative corpus
converges to oracle-level FPR and FDR by $K_c \approx 30$--50
(Figures~\ref{fig:corpus_fpr}--\ref{fig:corpus_fdr}).

The winner-selected corpus produces FPR $\approx 0.20$ and elevated
FDR at every corpus size, and collecting more data from the same
biased source does not help. The pooled-programs corpus produces
persistent FPR and FDR inflation that also does not converge to oracle
with more data
(Figures~\ref{fig:corpus_fpr}--\ref{fig:corpus_fdr}). Comparing
Figures~\ref{fig:corpus_avgn} and Table~\ref{tab:simB}, the winner
and pooled corpora stop faster than oracle for the same reason they
produce higher FPR. Their lower effective thresholds set a weaker
evidence standard that is inseparable from the weaker guarantees.

Together these results show that the need for explicit
multiple-testing correction across metrics is strongly coupled to EB
estimation quality. A well-calibrated prior absorbs most of the
inflation through the $(1-\hat{p})$ floor, while a miscalibrated
prior leaves the program back at the uncorrected baseline. BF-scale
robustness does not substitute for a correct null-rate estimate.

\subsection{When does the decision-theoretic advantage hold? (Setting~D)}
\label{sub:sim:shipping}

Settings~A--C evaluate all methods on error-rate metrics. Setting~D
asks the complementary question: how do the methods compare on the
decision-theoretic metric that expected-loss stopping optimizes for?
We define regret per decision as $|\delta|$ when the method makes the
wrong call and $0$ when it makes the right call, plus a shipping cost
$s \geq 0$ incurred every time a new variant is launched. The shipping
cost represents the non-metric cost of deploying a null-effect variant
(maintenance, codebase complexity, opportunity cost of the slot). When
$s = 0$, this reduces to the pure expected-loss criterion of
\citet{stucchio2015}. The DGP is the same normal model as Settings~A
and~B, with non-null effects drawn from $N(0, V^2)$, $V = 0.20$, null
rate $1 - p = 0.70$ (matching Settings~A--C), and $n_{\max} = 1{,}000$
per arm ($n_{\mathrm{sims}} = 5{,}000$, seed 99).

\begin{figure}[t]
  \centering
  \includegraphics[width=\linewidth]{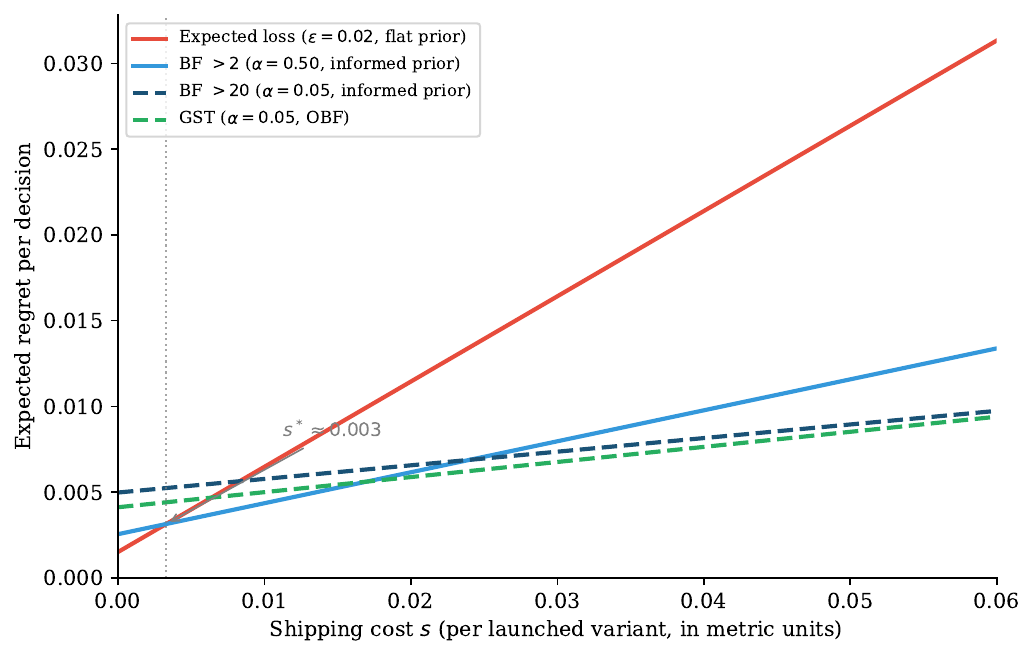}
  \caption{Setting~D: expected regret per decision as a function of
    shipping cost $s$ (null rate $1 - p = 0.70$, matching
    Settings~A--C). At $s = 0$, expected-loss stopping has the lowest
    regret. BF stopping at $\mathrm{BF} > 2$ (Ville bound
    $\alpha = 0.50$) overtakes it at $s^* \approx 0.003$. The slope
    of each curve reflects the method's ship rate.}
  \label{fig:shipping_cost}
\end{figure}

Figure~\ref{fig:shipping_cost} shows expected regret as a function of
$s$ for four configurations. At $s = 0$, expected-loss stopping has
the lowest regret because it stops fastest and its errors under $H_0$
are costless in metric units. However, its regret grows steeply with
$s$ because it ships the new variant in roughly half of all
experiments (ship rate $\approx 0.50$), including most null-effect
ones. BF stopping at its loosest calibration ($\mathrm{BF} > 2$,
Ville bound $\alpha = 0.50$) ships less often (ship rate
$\approx 0.18$) and overtakes expected-loss stopping at
$s^* \approx 0.003$, roughly $2\%$ of the mean absolute effect size
among non-null experiments. The strict configurations
($\mathrm{BF} > 20$ and GST at $\alpha = 0.05$) ship even less often
and dominate at higher shipping costs.

The figure makes the trade-off concrete. Expected-loss stopping is
optimal when deploying a null-effect variant is truly costless. Once
that cost exceeds a few percent of a typical real effect, methods that
require evidence before shipping achieve lower total regret.

\section{Discussion}
\label{sec:conclusion}

This paper organized Bayesian A/B testing configurations by their
 error-rate properties. We were somewhat surprised to find
how much of the recent Bayesian A/B testing literature is devoted to
achieving frequentist guarantees
\citep{johari2022,deng2016,deng2015,hagar2025,grunwald2024,schonbrodt2018,pawel2025}.
Naturally, every experimentation program produces
some FPR and FNR, whatever statistical framework drives its decisions. These are
operational properties of the program, and describe how the program behaves whether or not it optimizes for 
them. For cost functions with affine terminal payoffs, the optimal stopping rule
is often a BF threshold whose Ville bound directly yields a frequentist error rate.

Decision-theoretic formulations give a natural way to formalize what
the program cares about
\citep{wan2023,feit2019,goldberg2017,stucchio2015}. They can replace
conventional error rates or be used to derive them. The conventional
$\alpha = 0.05$ and $\beta = 0.20$ are rarely derived from the
program's actual costs. A cost function makes those assumptions
explicit. For some cost structures the result is a BF threshold
(\ref{app:bf_threshold}). For others, like expected-loss stopping,
something different falls out.

Specifying costs honestly is not trivial. Zero implementation cost is
a convenient assumption. For simple changes (reworded copy, reordered
list) it may hold. For new features, LLM-backed components, or
anything competing for app real estate, it is questionable. Even when
the implementation cost is a small fraction of the typical effect
size, methods that ship more conservatively achieve lower expected
regret than methods that ship without requiring statistical evidence
(Figure~\ref{fig:shipping_cost}).

Tier~3 delivers many of the properties that are often attributed to
Bayesian methods in general, but actually require a calibrated
empirical Bayes prior combined with BF stopping: automatic
multiplicity correction across metrics, a cure for the winner's curse
through calibrated shrinkage, and the lowest estimation error of any
configuration in our comparison. These do not come from using a
posterior or computing $P(B > A)$. They come from an EB prior fitted
to the program's historical effect-size distribution, paired with a
stopping rule whose error-rate guarantees survive optional stopping.
Without both, a Bayesian configuration has no more multiplicity
protection than a frequentist one. The EB prior, however, assumes that
future experiments are exchangeable draws from the same distribution
as past ones. This fails when effect sizes shrink over time as a
program exhausts its easiest wins, when different metrics have
systematically different effect-size distributions, or when the
program's scope shifts. A program with a stable, homogeneous testing
portfolio and hundreds of historical experiments can sustain a
calibrated prior. By contrast, a program where the landscape shifts between seasons,
where guardrail metrics and primary metrics live in different
effect-size regimes, or where the team is still learning what to test,
are better served by Tier~2 with explicit multiplicity corrections.

Within Tier~2 (BF stopping, GST, always-valid CS), the differences
are in stopping time, MSE, and whether an $n_{\max}$ commitment is
required. The MSE advantage of BF stopping over frequentist alternatives comes from the shrinkage embedded in the posterior mean estimator, not the stopping rule itself. Even a rough prior pulls extreme estimates toward plausible values, and this benefit is available at every tier. A frequentist can obtain similar benefits by applying shrinkage post hoc. Again, the choice of configuration is likely more important than the choice of framework.

Start with what risks your program needs to control. An organization
that needs to minimize expected cost per launch decision will arrive
at a different configuration than one that needs to bound the false
discovery rate across hundreds of experiments. The method should follow from
the risk, not the other way around. Every product decision made from
an A/B test carries risk. The configuration that best manages that risk
will depend on what the program needs to protect, not on whether the
method is called Bayesian or frequentist.

\section*{Acknowlegements}
The authors thank Sebastian Ankargren, Ryan Kessler, Lucas Vermeer, Heather Mathews, Edward Ronan, and Joseph Powers for feedback and suggestions to this paper. 

\pagebreak

\pagebreak
\appendix
\section{EB Shrinkage: Formal Details}
\label{app:eb}

\textbf{James--Stein result.} The James--Stein theorem \citep{james1961}
establishes that for $\hat{\delta} \sim N_d(\delta, \sigma^2 I)$ with
$d \geq 3$, the estimator
\begin{equation}
  \hat{\delta}_{\mathrm{JS}} = \left(1 - \frac{(d-2)\sigma^2}{\|\hat{\delta}\|^2}
  \right)\hat{\delta}
\end{equation}
dominates the MLE under squared error loss for all $\delta$. The result
assumes independent coordinates with equal variance; in practice,
metrics within an experiment have unequal variances and may be
correlated, so the analogy is approximate. No individual estimate is
guaranteed to improve. The guarantee is collective. The EB
posterior mean from Section~\ref{sec:tier3} is the empirical Bayes
instantiation of this result, with the shrinkage factor $A_i$ and posterior
weight $w_i$ playing the role of the adaptive multiplier.

\section{Sequential Testing: Formal Details}
\label{app:seq}

\textbf{E-variables and e-processes.} A nonneg\-ative random variable $E$ is an
e-variable for $H_0$ if $E_{P_0}[E] \leq 1$ for all $P_0 \in H_0$. A sequence
$(E_n)$ is an e-process if $E_\tau$ is an e-variable for every stopping time
$\tau$ \citep{grunwald2024,ramdas2023}. The BF sequence forms an e-process for
any proper prior because $E_{H_0}[\mathrm{BF}_n \mid X_1,\ldots,X_{n-1}] =
\mathrm{BF}_{n-1}$, so the BF is a nonnegative martingale under $H_0$
\citep{ville1939}.

\textbf{Hagar--Stevens joint design.} Under the normal-normal model with BF
stopping, define
\begin{align}
  \mathrm{FDR}(n, t) &= \frac{(1-p)\,P(\mathrm{BF}_n > t \mid H_0)}
    {(1-p)\,P(\mathrm{BF}_n > t \mid H_0) + p\,P(\mathrm{BF}_n > t \mid H_1)}, \\
  \mathrm{Power}(n, t) &= P(\mathrm{BF}_n > t \mid H_1).
\end{align}
The joint design problem finds $(n^*, t^*)$ to minimize $n$ subject to
$\mathrm{FDR}(n, t) \leq q$ and $\mathrm{Power}(n, t) \geq 1 - \beta$. Both
quantities have closed forms in terms of the noncentral chi-squared distribution
\citep{hagar2025}. The $\mathrm{FDR}(n, t)$ formula assumes $n$ is a fixed
sample size. It provides the correct design-time bound but does not bound
FDR when $n$ is replaced by a random stopping time $\tau$. The sequential pFDR
bound under BF thresholding follows the posterior-odds argument \citep{deng2016}.

\section{Type~M Errors in Bayesian Experimentation}
\label{app:typeM}

A Type~M error occurs when the magnitude of a statistically significant result
is inflated, with the effect real but the estimate at the threshold
substantially exceeding the true effect \citep{gelman2014}. The source is
selection. Conditioning on $\hat{\delta} > c$ selects for experiments where
noise pushed the estimate past the threshold, so the conditional mean of
$\hat{\delta}$ exceeds $\delta$. When power is low, this inflation can be
severe.

The EB posterior mean $\hat{\delta}_{\mathrm{EB}} = w_i \cdot A_i \cdot
\hat{\delta}$ is shrunk toward zero, partially correcting the Type~M bias
automatically and without conditioning on significance. The degree of
correction depends on prior calibration. A well-fitted prior provides
adaptive shrinkage that is heavier for noisy experiments. A miscalibrated
prior can leave the bias largely intact or, if the prior is biased upward,
make it worse. Power analysis via BFDA is necessary to bound the estimation
error conditional on stopping. \citet{schonbrodt2018} demonstrate that
underpowered BFDA designs produce upwardly biased estimates among
experiments that terminate before $n_{\max}$, with bias increasing as
power decreases.

\section{Simulation Design Details}
\label{app:sim}

\textbf{Setting~A.}
Normal DGP with $Y_{T,i} \sim N(\delta, 1)$, $Y_{C,i} \sim N(0, 1)$,
$\sigma = 1$ known. Hypotheses: $H_0\colon \delta = 0$,
$H_1\colon \delta = 0.2$ (Cohen's $d = 0.2$). Sequential monitoring every
50 obs per arm, with $n_{\max} = 5{,}000$ and $n_{\mathrm{sims}} = 10{,}000$.
Sufficient statistic: $\hat{\delta}_n = \bar{Y}_T - \bar{Y}_C \sim
N(\delta, 2/n)$. Prior families: flat $\pi(\delta) \propto 1$,
weakly informative $N(0, V^2)$ with $V = 0.4 = 2\delta_{\mathrm{MDE}}$,
and matched Gaussian $N(0, V^2)$ with $V = 0.2 = \delta_{\mathrm{MDE}}/\sigma$.
Expected-loss threshold $\varepsilon = 0.02$, a commonly used setting
\citep{stucchio2015}, included to illustrate the contrast in
Sections~\ref{sec:model} and~\ref{sub:decision_theoretic}.

BF closed form (specializing Eq.~\ref{eq:bf_closed} to $\sigma = 1$,
$\sigma^2_{\mathrm{obs}} = 2/n$):
$\Lambda_n = \sqrt{\sigma^2_{\mathrm{obs}} / (\sigma^2_{\mathrm{obs}}+V^2)}
\cdot \exp(\hat\delta_n^2 V^2 /
(2\sigma^2_{\mathrm{obs}}(\sigma^2_{\mathrm{obs}}+V^2)))$.
Expected-loss computation uses $E[\max(\delta,0)] = \mu_n\Phi(\mu_n/s_n)
+ s_n\varphi(\mu_n/s_n)$ from the $N(\mu_n, s_n^2)$ posterior, where $s_n$
is the posterior standard deviation.

\textbf{Setting~B: sequential methods comparison.}
Common DGP, same normal model as Setting~A with $n_{\max} = 1{,}000$ per arm,
$n_{\mathrm{sims}} = 5{,}000$, seed 99. Program-level non-null rate $p = 0.30$ (null rate $1-p = 0.70$).
Monitoring cadence is method-specific. In this simulation, GST uses
14 or 21 looks (Lan--DeMets $\alpha$-spending allows flexible look
times within the committed $n_{\max}$). All other methods (WS--R,
BF/Gaussian, BF/EB) monitor after every observation pair.

GST (O'Brien--Fleming) uses two schedules approximating daily monitoring of
a production experiment, 14 looks (2-week run) and 21 looks (3-week run),
equally spaced in information time. Boundaries are computed by forward
density recursion under one-sided OBF $\alpha$-spending
$f(t) = 1 - \Phi(z_{1-\alpha}/\sqrt{t})$ \citep{landemets1983}. Cumulative
Type~I error is verified to $\le 0.05$ by direct simulation under $H_0$.

Always-valid CS (Waudby-Smith--Ramdas) uses the mixture-martingale confidence
sequence \citep{waudby2024} with $\rho^2 = \delta_{\mathrm{MDE}}^2$ as the
mixing variance, tuned to concentrate power at the MDE. We stop when the CS
excludes zero, and report the CS at stopping.

BF configurations: (i) Gaussian with $V = \delta_{\mathrm{MDE}}/\sigma = 0.2$,
(ii) oracle EB with true $(p = 0.30, V = 0.20)$, using the mixture e-process
$\Lambda_n = (1-p) + p \cdot \Lambda^N_n(V)$, (iii) winner-selected EB
($\hat{p} \approx 1.0$, $\hat{V}\approx 0.23$ from a corpus of only
$z > 1.645$ experiments), and (iv) pooled-programs EB ($\hat{p} \approx 0.51$,
$\hat{V} \approx 0.37$ from mixing with a second program at $p = 0.80$,
$V = 0.40$).

Point estimates are the posterior mean for Bayesian methods and
$\hat\delta$ at stopping for GST and always-valid CS. MSE is computed among
$H_1$ experiments that stopped. Interval width uses the posterior credible
interval for Bayesian methods, the confidence sequence width for
always-valid CS, and the stagewise CI width for GST.

\textbf{Setting~C.}
Experiment-level all-or-nothing DGP with $K = 10$ metrics. Each experiment
is fully null (all metrics $\delta = 0$) with probability $1 - p = 0.70$
or fully non-null (all metrics $\delta \sim N(0, V^2)$) with probability
$p = 0.30$, and true $V = 0.20$. The corpus is a flat list of per-metric
historical observations from a single program, treated as exchangeable
draws from one shared mixture, and one $(\hat{p}, \hat{V})$ is fitted and
applied to every metric in the test experiment. Under the all-or-nothing
DGP the per-metric non-null rate $p$ equals the experiment-level non-null
rate $p$, so the same corpus-derived $\hat{p}$ encodes both.
Corpus sizes $K_c \in \{10, 30, 50, 100, 500\}$, with corpus observations
at $n = 400$ per arm ($\sigma_c^2 = 2/400 = 0.005$). The EB prior is
fitted by MLE (L-BFGS-B, 5 restarts). BF stopping uses the mixture
e-process $\Lambda_n = (1-\hat{p}) + \hat{p}\cdot\Lambda^N_n(\hat{V}) > 20$.
Corpus conditions: Oracle EB (true $p$, $V$ known), representative corpus,
winner-selected corpus (only $z > 1.645$ kept), and pooled programs (test
program mixed with Program~C, $p = 0.80$, $V = 0.40$, evaluated on
Program~B, $p = 0.30$, $V = 0.20$). $M = 15$ corpus draws per $K_c$,
$n_{\mathrm{sims}} = 2{,}000$, seed 77.

\begin{table*}[t]
\centering
\caption{BF prior calibration sensitivity (supplementary to Setting~A).
  BF stopping ($\Lambda_n > 20$) under varying $V$. The FPR bound is invariant to
  $V$ miscalibration, and only power efficiency varies. Normal DGP,
  $n_{\max} = 5{,}000$.
  Pwr@$n_{\mathrm{fix}}$ is fixed-sample power at $n = n_{\max}$
  without sequential stopping.}
\label{tab:bfsens}
\begin{tabular}{llrrr}
\toprule
Prior & $V$ & FPR & Pwr@$n_{\mathrm{fix}}$ & Avg.\ $n$ \\
\midrule
Oracle      & 0.20 ($=\delta_{\mathrm{MDE}}$)          & 0.019 & 0.473 & 502 \\
Optimistic  & 0.40 ($=2\delta_{\mathrm{MDE}}$)         & 0.016 & 0.450 & 525 \\
Pessimistic & 0.10 ($=0.5\delta_{\mathrm{MDE}}$)       & 0.019 & 0.354 & 560 \\
Winner-sel. & 0.35                             & 0.017 & 0.458 & 518 \\
\bottomrule
\end{tabular}
\end{table*}

\section{Bellman Derivation of the BF Threshold Rule}
\label{app:bellman}

Let $\pi_0, \pi_1 = 1-\pi_0$ be the prior probabilities of $H_0, H_1$.
At each $n$ the decision maker observes $\mathcal{F}_n = \sigma(Y_{1:n})$
and picks one of three actions: stop and accept $H_0$ (terminal decision
$D_\tau = 0$), stop and reject $H_0$ ($D_\tau = 1$), or continue
(requiring $c > 0$ for the continuation cost to penalize delay). On a
path where the true model is $M \in \{H_0, H_1\}$, the loss is
\begin{equation}
  \label{eq:loss}
  \ell(\tau, D_\tau, M) = c\,\tau
    + K_I\,\mathbf{1}\{D_\tau{=}1,\,M{=}H_0\}
    + K_{II}\,\mathbf{1}\{D_\tau{=}0,\,M{=}H_1\}.
\end{equation}
The Bayes risk is
\begin{equation}
  \label{eq:bayes_risk}
  R(\tau, D_\tau) = \pi_0\bigl[c\,\mathbb{E}_0[\tau] + K_I\,\alpha\bigr]
    + \pi_1\bigl[c\,\mathbb{E}_1[\tau] + K_{II}\,\beta\bigr],
\end{equation}
where $\alpha = \mathbb{P}_0(D_\tau{=}1)$ and
$\beta = \mathbb{P}_1(D_\tau{=}0)$ are the FPR and FNR of the rule.

\textbf{Posterior summary.} The posterior probability
$\pi_n := \mathbb{P}(H_1 \mid \mathcal{F}_n)$ is related to the Bayes
factor through
\begin{equation}
  \pi_n = \frac{\pi_1\,\mathrm{BF}_{10}(n)}
    {\pi_0 + \pi_1\,\mathrm{BF}_{10}(n)},
  \qquad
  \mathrm{BF}_{10}(n) = \frac{\pi_0}{\pi_1}\,\frac{\pi_n}{1-\pi_n},
\end{equation}
a smooth monotone bijection. The losses for the two terminal actions are
linear in $\pi_n$: accepting $H_0$ costs $K_{II}\,\pi_n$, rejecting $H_0$
costs $K_I\,(1-\pi_n)$, and continuing costs
$c + \mathbb{E}[V_{n+1}(\pi_{n+1}) \mid \pi_n]$.

\textbf{Bellman equation.} Define the value function as
$V_n(\pi) := \inf_{(\tau,D)} \mathbb{E}[\ell(\tau,D,M) - c\,n \mid
\pi_n = \pi]$. The optimality equation is
\begin{equation}
  V_n(\pi) = \min\bigl\{K_{II}\,\pi,\;\,K_I\,(1-\pi),\;\,
    c + \mathbb{E}[V_{n+1}(\pi_{n+1}) \mid \pi_n = \pi]\bigr\}.
\end{equation}
By induction \citep{arrow1949}, $V_n$ is concave in $\pi$ on $[0,1]$.
For any fixed policy, the expected cost is linear in $\pi$ (it is a
$\pi$-weighted mixture of costs under $H_0$ and $H_1$). The infimum
of a family of linear functions is concave, so $V_n$ is concave. Since $V_n$ is concave and the terminal costs are linear, the
set of $\pi$ for which continuing is strictly optimal is
an open interval $(p_*, p^*) \subseteq [0,1]$, with $p_* \leq \pi^\dagger
\leq p^*$ where $\pi^\dagger = K_I/(K_I+K_{II})$ is the intersection of
the two terminal-cost lines.

\textbf{Optimal rule in terms of BF$_{10}$.} Writing the rule in terms of
$\mathrm{BF}_{10}$ via the bijection above, the optimal policy is:
\begin{equation}
  \tau^* = \inf\{n : \mathrm{BF}_{10}(n) \notin (B^*, A^*)\},
  \quad
  D_{\tau^*} = \mathbf{1}\{\mathrm{BF}_{10}(\tau^*) \geq A^*\},
\end{equation}
with thresholds
\begin{equation}
  A^* = \frac{\pi_0}{\pi_1}\,\frac{p^*}{1-p^*}, \qquad
  B^* = \frac{\pi_0}{\pi_1}\,\frac{p_*}{1-p_*}.
\end{equation}
The boundary points $(p_*, p^*)$ are determined by the smoothing
(smooth-pasting) conditions of the optimal stopping problem and have no
general closed form. Wald's approximation \citep{wald1948} give
$A^* \approx (1-\beta)/\alpha$ and $B^* \approx \beta/(1-\alpha)$ as
$\alpha, \beta \to 0$.

\begin{remark}[Cost-to-rate mapping]
\label{rem:cost_rate}
Setting $A = 1/\alpha$ and $B = \beta$ in the threshold form above
produces a rule with anytime FPR~$\leq \alpha$ (Ville on $\mathrm{BF}_{10}$)
and prior-averaged FNR~$\leq \beta$ (Ville on $1/\mathrm{BF}_{10}$;
\ref{app:fnr_bound}). The mapping is approximately invertible: a platform
specifying $(\alpha, \beta)$ implicitly operates at cost ratios
$(K_I/c, K_{II}/c)$ that rationalize those rates, and a platform
specifying $(c, K_I, K_{II})$ inherits calibrated $(\alpha^*, \beta^*)$
via the Wald--Pollak approximation. The Ville-calibrated thresholds
$(1/\alpha, \beta)$ are more conservative than the cost-optimal
$((1-\beta)/\alpha,\; \beta/(1-\alpha))$ by factors $1/(1-\beta)$ and
$(1-\alpha)$ respectively. At $\alpha = 0.05$, $\beta = 0.20$ the upper
threshold is inflated by 25\%, a modest price for the clean Ville
guarantee.
\end{remark}

\section{Prior-Averaged FNR Bound}
\label{app:fnr_bound}

For any proper prior $\pi$ under $H_1$, the inverse process
$\{1/\mathrm{BF}_{10}(n)\}$ is a nonnegative martingale under the
prior-predictive distribution
$\mathbb{P}_\pi(\cdot) = \int P(\cdot \mid \delta)\,d\pi(\delta)$, with
$\mathbb{E}_\pi[1/\mathrm{BF}_{10}(n)] = 1$ \citep{ville1939}. Ville's
inequality gives, for any stopping time $\tau$ and any $B \in (0,1]$,
\begin{equation}
  \label{eq:ville_inverse}
  \mathbb{P}_\pi\!\Bigl(\inf_n \mathrm{BF}_{10}(n) \leq B\Bigr)
  \;\leq\; B.
\end{equation}
Under the two-threshold rule of Section~\ref{sub:decision_theoretic},
accepting $H_0$ ($D_\tau = 0$) requires
$\mathrm{BF}_{10}(\tau) \leq B$, so
$\mathbb{P}_\pi(D_\tau = 0) \leq B$. Setting $B = \beta$ bounds the
prior-averaged FNR at $\beta$, provided accepting $H_0$ requires
actually crossing the lower boundary. Paths that reach $n_{\max}$
without crossing either threshold are additional false negatives not
covered by this bound. For large $n_{\max}$ under $H_1$ the
probability of non-crossing is negligible (the BF diverges), but the
bound as stated applies only to the crossing event.

The FPR bound is unconditional: it holds for any true $\delta$ under
$H_0$ (a point hypothesis). The FNR bound is averaged over
$\delta \sim \pi$: for specific values of $\delta$, especially small
effects near zero, the conditional FNR may exceed $\beta$.

\section{BF Stopping Implies CI Exclusion}
\label{app:bf_ci}

Under the conjugate model with prior $\delta \sim N(0, V^2)$ and data
precision $\kappa_n = N_{E,n}\,V^2/\sigma^2$, the BF is
(Eq.~\ref{eq:bf_closed})
\[
  \mathrm{BF}_{10}(n)
  = (1+\kappa_n)^{-1/2}\,
    \exp\!\Bigl(\frac{Z_n^2\,\kappa_n}{2(1+\kappa_n)}\Bigr).
\]
Solving $\mathrm{BF}_{10}(n) \geq 1/\alpha$ for $Z_n^2$ gives
\begin{equation}
  \label{eq:bf_z_threshold}
  Z_n^2 \;\geq\; \frac{1+\kappa_n}{\kappa_n}
    \bigl[\log(1+\kappa_n) + 2\log(1/\alpha)\bigr].
\end{equation}

The posterior is $\delta \mid \text{data} \sim N(\hat\mu_n, v_n)$ with
$\hat\mu_n = \hat\delta_n\,\kappa_n/(1+\kappa_n)$ and
$v_n = V^2/(1+\kappa_n)$. The $(1-\alpha)$ equal-tailed
credible interval excludes zero iff
\[
  \hat\mu_n^2 / v_n > z_{\alpha/2}^2,
\]
which in terms of $Z_n$ reduces to
\begin{equation}
  \label{eq:ci_z_threshold}
  Z_n^2 \;>\; z_{\alpha/2}^2\,\frac{1+\kappa_n}{\kappa_n}.
\end{equation}

Comparing~\eqref{eq:bf_z_threshold} and~\eqref{eq:ci_z_threshold}, the BF
threshold exceeds the CI threshold whenever
$\log(1+\kappa_n) + 2\log(1/\alpha) > z_{\alpha/2}^2$. Since
$\kappa_n > 0$, it suffices to verify $2\log(1/\alpha) > z_{\alpha/2}^2$.
Define $f(\alpha) = 2\log(1/\alpha) - z_{\alpha/2}^2$. We show
$f > 0$ on $(0,1)$ by proving $f$ is strictly decreasing with
$\lim_{\alpha \to 1^-} f(\alpha) = 0^+$. Differentiating,
$f'(\alpha) = -2/\alpha + z_{\alpha/2}/\varphi(z_{\alpha/2})$, where
$\varphi$ is the standard normal density. The Gaussian tail bound
gives $z/\varphi(z) < 1/(1-\Phi(z)) = 2/\alpha$ for all $z > 0$, so
$f'(\alpha) < 0$ on $(0,1)$. Since $z_{\alpha/2} \to 0$ and
$\log(1/\alpha) \to 0$ as $\alpha \to 1^-$, with $2\log(1/\alpha)$
dominating $z_{\alpha/2}^2$, we have $f(\alpha) \to 0^+$. A strictly
decreasing function with limit $0^+$ is strictly positive.
Therefore \eqref{eq:bf_z_threshold} implies~\eqref{eq:ci_z_threshold},
and $\mathrm{BF}_{10}(n) \geq 1/\alpha$ implies the credible interval
excludes zero.\qed

\section{BF Threshold Structure for General Cost Functions}
\label{app:bf_threshold}

The following result shows that the BF threshold structure is not
specific to the misclassification-cost loss but holds for a large class
of sequential decision problems.

Consider a sequential three-action problem with finite horizon
$T \geq 1$, where at each $n$ the decision maker chooses to continue,
accept $H_0$, or reject $H_0$. Let $\pi_n = P(H_1 \mid Y_{1:n})$
under a proper prior with $P(H_0), P(H_1) \in (0,1)$. Assume: (i)~the
expected costs of all three actions are affine in $\pi_n$ at each
fixed $n$, (ii)~$\pi_n$ is a sufficient statistic for the decision,
and (iii)~for any concave $V$, the map
$\pi \mapsto \mathbb{E}[V(\pi_{n+1}) \mid \pi_n = \pi]$ is concave.
Then:

\begin{enumerate}
\item[(a)] The value function $V_n(\pi)$ is concave in $\pi$ for every $n$.
\item[(b)] The optimal continuation region at each $n$ is an interval
  $(p_*^n, p^{*n})$ in $[0,1]$.
\item[(c)] Via the monotone bijection $\pi \leftrightarrow
  \mathrm{BF}_{10}$, the optimal policy is a threshold rule on the
  Bayes factor. The thresholds generally depend on $n$.
\item[(d)] For any constant $C > 0$ and any stopping rule that rejects
  $H_0$ only when $\mathrm{BF}_{10} \geq C$,
  $P(\text{reject } H_0 \mid H_0) \leq 1/C$. Setting $C = 1/\alpha$
  gives $\mathrm{FPR} \leq \alpha$ regardless of the loss function.
\end{enumerate}

\textit{Proof.}
(a)~For any fixed policy (a rule specifying when to stop and which
terminal action to take), the expected cost from time $n$ onward is
affine in $\pi_n$: by the tower property, it is a $\pi_n$-weighted
mixture of expected costs under $H_0$ and $H_1$, each of which is a
constant determined by the policy. $V_n$ is the infimum over all such
policies, and the pointwise infimum of a family of affine functions is
concave.

(b)~Define $h_i(\pi) = f_i(n,\pi) - V_n(\pi)$ for each terminal action
$i \in \{0,1\}$. Each $h_i$ is convex (affine minus concave) and
nonnegative ($V_n \leq f_i$ by construction). The zero set of a
nonnegative convex function is a closed convex set, hence a closed
interval on $[0,1]$. The accept-$H_0$ region is
$[0, p_*^n]$ and the reject-$H_0$ region is $[p^{*n}, 1]$, giving the
continuation region as the interval between them.

(c)~The map $\mathrm{BF}_{10} = (\pi_0/\pi_1)\,\pi/(1-\pi)$ is
strictly increasing, so it maps intervals to intervals.

(d)~Under $H_0$, $\{\mathrm{BF}_{10}(n)\}$ is a nonnegative martingale
with $\mathrm{BF}_{10}(0) = 1$. Ville's inequality gives
$P(\sup_n \mathrm{BF}_{10}(n) \geq C \mid H_0) \leq 1/C$. This is a
property of the likelihood ratio and requires only a proper prior, not
any assumption about the loss function. \qed

\textbf{Example.} Under the two-point model, any cost determined by
which hypothesis is true is automatically affine in $\pi_n$: if
rejecting $H_0$ costs $K_I$ when $H_0$ is true and nothing otherwise,
the expected cost is $K_I(1 - \pi_n)$; accepting $H_0$ costs
$K_{II}\,\pi_n$. More generally,
$\mathbb{E}[g(\delta) \mid \text{data}] =
g(0)(1-\pi_n) + g(\delta_1)\pi_n$ for any function $g$, so the
condition is mild. Under continuous priors (see below), the full
posterior is no longer determined by a single probability, and costs
involving nonlinear posterior functionals (such as the truncated
expectation in the expected opportunity loss rule) break the affine
structure.

The affine-cost condition (i) holds whenever the sampling and
terminal costs depend on the unknown $\delta$ only through
$\mathbb{E}[\delta \mid \text{data}]$, which is affine in $\pi_n$
under the two-point model ($\delta \in \{0, \delta_1\}$). We have verified numerically that the theorem applies to the utility
function of \citet{wan2023}, which includes customer impact during the
experiment, time-decaying launch benefits, and hurdle costs.
\citeauthor{wan2023} discretize the experiment into $T$ observation
windows, each incurring a sampling cost. Under representative
parameters ($T = 10$ observation windows, post-experiment horizon
$H = 52$, treatment group $N_{\mathrm{Tr}} = 1{,}000$, total
population $N = 2{,}000$, sampling cost $c = 1$ per observation window, hurdle cost
$c_h = 50$, effect size $\delta_1 = 0.1$), backward induction on a
2001-point posterior grid yields an upper threshold
$A^* \approx 0.42$ in BF space (well below $1$, reflecting that the cost structure favors launching even under weak evidence), roughly constant through observation window~6 and
declining toward the deadline as the option value of waiting shrinks.
The BF threshold structure holds at every time step.
The mapping from hurdle cost to threshold is approximately
$A^* \approx (c_h - c)/(\delta_1 N_{\mathrm{Ctrl}} + c)$
in a one-step lookahead approximation, giving
$\alpha_{\mathrm{eff}} \approx (\delta_1 N_{\mathrm{Ctrl}} + c)/(c_h - c)$.
As $c_h \to 0$ the continuation region vanishes (it is optimal to ship
immediately without experimentation), and $A^* = 1$ (equivalent to
$\alpha = 1$) requires $c_h \approx \delta_1 N_{\mathrm{Ctrl}} + c$.

\textbf{Extension to continuous priors.} The proposition above is
stated for the two-point model, where $\pi_n = P(H_1 \mid
\mathcal{F}_n)$ is the state variable. The configurations in Sections
3--4 use a continuous prior $\delta \sim N(0, V^2)$, under which
$\pi_n$ is not defined. The Bellman argument extends with the
posterior mean $\mu_n = E[\delta \mid \mathcal{F}_n]$ as the state
variable. Under the conjugate normal model, $\mu_n$ is a martingale
and the conditional distribution $\mu_{n+1} \mid \mu_n$ is Gaussian
with deterministic variance, so concavity preservation holds by the
standard Gaussian convolution argument
\citep[Ch.~4]{shiryaev1978}. The Bellman equation yields interval
continuation regions $(m_{\mathrm{low}}(n),\, m_{\mathrm{high}}(n))$
in $\mu_n$-space.

Under the symmetric $N(0, V^2)$ prior, $\mathrm{BF}_{10}$ depends on
$\mu_n^2$ (equivalently $Z_n^2$) and is therefore monotone in
$|\mu_n|$, not in $\mu_n$. For the non-directional detection problem
with symmetric costs ($K_I = K_{II}$), the value function inherits the
symmetry $V_n(\mu) = V_n(-\mu)$, forcing
$m_{\mathrm{high}} = -m_{\mathrm{low}}$. A symmetric interval in
$\mu_n$ maps to an interval in $|\mu_n|$ and hence to a BF threshold.
The optimal policy is then a BF threshold rule with a sign check for
the directional decision.

For asymmetric costs ($K_I \neq K_{II}$), the optimal thresholds
satisfy $m_{\mathrm{high}} \neq -m_{\mathrm{low}}$ and the single BF
threshold is a constrained approximation to the optimal policy. Since
the symmetric constraint is not binding at $K_I = K_{II}$, we expect
the gap to be small for moderate asymmetry, though we have not
quantified it.

Part~(d) (Ville FPR bound) requires only a proper prior and applies
without modification. With the symmetric BF at threshold $1/\alpha$
and a sign check, the directional FPR is at most $\alpha/2$.

\end{document}